Abstract
The recent development of the lattice gas automata method and its
extension to the lattice Boltzmann method have provided new
computational schemes for solving a variety of partial differential equations
and modeling chemically reacting systems. The lattice gas method,
regarded as the simplest microscopic and kinetic approach which
generates meaningful macroscopic dynamics, is fully parallel
and can, as a result,  be easily programmed on parallel machines.
In this paper,  we introduce the basic principles of the lattice
gas  method and the lattice Boltzmann method, their numerical implementations
and applications to   chemically reacting systems. Comparisons of
the lattice Boltzmann method with the lattice gas technique and other
traditional numerical schemes, including the finite difference scheme
and the pseudo-spectral method, for solving the Navier-Stokes
hydrodynamic fluid flows will be discussed.  Recent developments
of the lattice gas and the lattice Boltzmann method and their applications to
 pattern formation in chemical reaction-diffusion systems, multiphase fluid
flows and polymeric dynamics will be presented.
\\
\newcommand {\be}{\begin{equation}}
\newcommand {\ee}{\end{equation}}
\newcommand {\bea}{\begin{eqnarray}}
\newcommand {\eea}{\end{eqnarray}}

\def\br{{\bf r}}

\def\bbrel#1#2#3{\mathrel{\mathop{\kern0pt #2}\limits^{#1}_{#3}}}

\documentstyle [11pt] {article}
\begin{document}
\large
\title{\bf Lattice Methods and Their Applications To Reacting Systems}
\author{S. Chen, S. P. Dawson, G. D. Doolen, D. R. Janecky\thanks{Isotope and
Nuclear Chemistry Division, Los Alamos National Laboratory, Los Alamos, NM
87545} and A. Lawniczak\thanks{Department of Mathematics and Statistics,
University of Guelph, Guelph, Ont N1G2W1, Canada}\\
  [.5cm]{ Theoretical Division}\\
        { and}\\
        {Center for Nonlinear Studies }\\
        {Los Alamos National Laboratory}\\
        {Los Alamos, NM 87545}\\[.4cm]}

\maketitle
\date{}
\begin{center}
{\bf Abstract}
\end{center}
The recent development of the lattice gas automata method and its extension to
the lattice Boltzmann method have provided new computational schemes
for solving a variety of partial differential equations and modeling
chemically reacting systems. The lattice gas method, regarded as the simplest
microscopic and kinetic approach which generates meaningful macroscopic
dynamics, is fully parallel and can, as a result,
 be easily programmed on parallel
machines.  In this paper,  we introduce the basic principles of the
lattice
gas  method and the lattice Boltzmann method, their
 numerical implementations and
 applications to   chemically reacting systems.
Comparisons of
the lattice Boltzmann method with the lattice gas technique and other
traditional numerical schemes, including the finite difference scheme
and the pseudo-spectral
method, for solving the Navier-Stokes hydrodynamic fluid flows will
 be discussed.  Recent developments of the lattice gas and the lattice
Boltzmann method and their applications to
 pattern formation in chemical reaction-diffusion systems,  multiphase fluid
flows and polymeric dynamics
will be presented.

\vspace{0.3in}

\pagebreak
\section{Introduction}

Lattice methods, including the lattice gas automaton
method\cite{fhp1,wolf,fhp2,gary0,gary1,gary2,syc0} and
 its derivative, the lattice Boltzmann
 method\cite{macn-zan,hig-jim,ccmm,qian}, are
powerful alternatives to traditional numerical schemes for solving partial
 differential equations and modeling physical systems, particularly for
 simulating fluid flows whose dynamics are described by the Navier-Stokes
 equations. In finite difference and finite element methods, the
 given macroscopic equation  is solved
by some specific numerical discretization while in
 full molecular dynamics descriptions,  each individual particle is
closely and accurately followed.
Unlike these traditional numerical schemes, the idea behind most
 lattice methods is to construct a
simplified molecular dynamics that incorporates many of the essential
 characteristics of the real microscopic processes
so that the macroscopic averaged properties
obey the desired
macroscopic equations.
The lattice method  can be described as  a mesoscopic model
which occupies a position between the full molecular
dynamics  and macroscopic descriptions by means of
partial differential equations.

Even though lattice models originate from a particle picture,
 they are
mainly
focused on averaged macroscopic behavior. In
 most cases lattice methods are developed as
numerical schemes rather than microscopic models for understanding fundamental
physics. However,
the utilization of the  particle description or
 the kinetic equation
provides some of the advantages of
 molecular dynamics, including clear physical
pictures, easy implementation of boundaries and fully
parallel algorithms.

A common feature of the lattice gas automaton method and the lattice
 Boltzmann method is the
 discretization of
space and time.   Lattice methods are purely local, parallel and very fast.
A speed of $10^{11}$ cell updates per second on a
Connection-Machine 5 can be obtained for a simple lattice gas model.
The simulation speed of the three-dimensional lattice Boltzmann
model is about 2 times faster than the pseudo-spectral method
for simulating periodic geometry
for the same spatial resolution. An important property of lattice
methods is that simulations for flow in simple and complex
geometries have the same speed and  the same efficiency, while all other
methods, including the spectral method, are unable to model
complicated geometries efficiently. This makes the lattice methods more
suitable for simulating flows in complicated geometry, such as porous
media flows. Usually code development  for lattice methods is considerably
faster and easier than other traditional schemes.

The development of a lattice model includes
the following three steps: (i) the construction of the model,
(including the design of a particle collision operator and
the choice of conservation rules)
(ii) the derivation of the macroscopic dynamics; for most cases, this
step involves a Chapman-Enskog expansion or a multiscaling
analysis;  (iii) the numerical
validation of the theory and
application of the model to real situations.
For a given macroscopic equation,
the design of the corresponding microscopic lattice method is an
inverse problem.  There are no general principles to guarantee
that one can find such a
lattice gas model.
Nevertheless, many lattice models already exist for solving partial
differential equations, including the
Navier-Stokes
 equations\cite{fhp1,fhp2,clav,hass,kadanoff,wolf,phy-report,sycjsp},
Burgers equation\cite{bogh,cheng},
the  Poisson equation\cite{hchen4}, the wave
equation\cite{hchen1,huang},
the diffusion equation\cite{binder0,binder1,ladd,qian1};
and for modeling physical phenomena, including
flow through porous media\cite{bala,roth1,schen1},
turbulent flows\cite{succi2,loh,succi3,sycjsp,daniel1,McNphys},
phase transitions\cite{hchen0,syc5,app0},
multiphase flows\cite{bur,roth2,somers,schen3,gun,daryl},
chemically-reacting flows\cite{dab,refkap2,sear,refkap1,refsil},
thermohydrodynamics\cite{burg,schen2,chopard0,molvig,frank1,scalo},
magnetohydrodynamics\cite{mon1,hudong-bill,ccmm},
liquid crystals\cite{syc-D,liu} and
semiconductors\cite{rie,moria}. In this review, we will
focus  our attention to applications related to
chemically reacting fluid flows,  even though applications of the
lattice methods are not limited to problems of fluids.

It is natural to use lattice methods for
simulating compressible
fluids because of the intrinsic compressibility
in the particle representation.
Most models developed so far, however, are mainly
 focused on the incompressible
limit of the lattice methods. Recent numerical simulations have demonstrated
that the solution of
the lattice methods  converges
 to the solution of the Navier-Stokes equations
for flow velocities much smaller than the sound speed\cite{reider}

Theoretical studies of lattice systems are also
relatively simpler than classical particle
systems.
Transport properties\cite{cor,luo,frank,henon,kong,zanetti,ernst2}
 and long time
behavior of correlations\cite{ernst0,ernst1} can often
 be exactly solved, while they are more
difficult to obtain
for continuum systems.

Most two-dimensional lattice gas models have been
 based on the hexagonal lattice. Three-dimensional
lattice gas models are based on the Face-Centred-Hypercubic lattice\cite{ddh1}
dictated by the isotropy requirements for
 the stress tensor and convection term in the Navier-Stokes equations.
Lattice Boltzmann methods can be
adapted to many different lattices, such
as the
9-directional square lattice\cite{qian,daniel1} in two dimensions and the
body-centered
cubic lattice in three dimensions\cite{sycjsp,qian}. In order to focus on the
fundamental
principles of the lattice methods, we will only discuss
 dynamics based on the hexagonal lattice. Some three-dimensional simulation
results will be presented
in the following sections.

\section{Lattice Gas and Lattice Boltzmann Methods}

\subsection{Principles of the Lattice Gas Method}

Unlike the molecular dynamics and
discrete velocity models\cite{broad,caban},  lattice gas automata
methods
emphasize  discreteness and
use only very few velocity values.
Lattice gas methods for hydrodynamic systems were
originally proposed by
Hardy, de Pazzis and Pomeau (HPP)\cite{hardy,hardy1} on a square lattice in two
dimensions for understanding transport
properties in hydrodynamics.
Serious problems in the square lattice include
spurious  momentum
conservation and anisotropy of the stress tensor.
The important contribution  made
 by Frisch, Hasslacher
and Pomeau (FHP)\cite{fhp1} was to show that a hexagonal lattice
gas model
 has an isotropic stress tensor and
 almost eliminates the spuriously
 conserved quantities.

Lattice gas methods locally
conserve
total mass, momentum and energy exactly,
with no round-off error, therefore these algorithms
are
unconditionally
stable.   Moreover,
lattice gas methods can easily handle very
complicated geometry and boundary conditions. For
example, the nonslip condition
can be implemented by particle velocity reversal
at the boundaries.

There are two main limitations for lattice gas models in
the simulation of fluid flows.
 First, the Reynolds numbers
 and
Mach numbers can not be
too large. On existing large computers, Reynolds numbers
up to 10,000 can be modelled in two dimensions
and up to 1000 in three dimensions.
The incompressible Navier-Stokes
equations are derivable only in the
low Mach number limit\cite{majda}. This
restriction can be relaxed if one uses
multiple discrete velocity models\cite{broad,caban}.

Second, local quantities
in lattice gas models are usually
noisy\cite{mong2,schen7}, requiring spatial and time averaging
in order to study the macroscopic dynamics.
 This requires large computer
memory or long
computational time. On the other hand,   fluctuations
are important for studying several physical details,
 such as the noise-driven pattern formation.
The lattice gas method is particularly suitable for these purposes.
 Lattice Boltzmann methods, in which real numbers replace the
bits in lattice gas model, greatly reduce the fluctuations.

Furthermore, there are two unsolved problems when one uses the lattice gas
method for simulating the Navier-Stokes fluids: the non-Galilean invariance
of convection and the velocity dependence of the equation of state\cite{fhp2}.
Both problems are effects of  the restriction to
 a finite number of velocities
and the Fermi-statistics employed to minimize memory
requirements.

The basic two-dimensional  FHP lattice gas model consists of
 identical particles moving on a hexagonal
lattice.  The
lattice spacing, $c$, is unity and
 all particles have the same mass, momentum magnitude and
kinetic energy, except for rest particles which are often included in order
to increase the Reynolds number range. There are six nonzero momentum states
associated with the
 directions to the nearest neighbors. An exclusion principle
is usually imposed which requires that no more than one particle
at a given site can have the same momentum. This exclusion rule is
included in order to minimize memory requirements, allowing bit operators.
Let
$N_i({\bf x},t)$ $(i=0, 1,\ldots ,6)$ be  the particle occupation in
state $i$ at site $\bf x$ and time $t$, then $N_i = 1$ or $0$ represents
a particle existence in  the $i$ state.  Similar to classical molecular
dynamics, there are two
microscopic updating processes at each discrete time step:
streaming (particle moving)
and  particle collision. In the streaming process, particles move to the
 nearest neighbor sites along their momentum directions.
In the collision process, particles at each
site are scattered in such a way that the total particle number
($= \sum_{i=0}^6 N_i$)
and the total momentum ($= \sum_{i=0}^6 {\bf e}_i N_i$) are conserved
at each site, where ${\bf e}_i (i=1,\cdot\cdot\cdot,6)$ is the unit vector
pointing to the nearest neighbors and ${\bf e}_0 = 0$.
The kinetic equation for the particle state occupation can
 be written in a simple form,

\begin{equation}
N_i({\bf x} + {\bf e}_i,t+1) = N_i({\bf x}, t) + \Omega_i,
\end{equation}
where $\Omega_i$ is the collision operator, which includes
the creation or annihilation of a particle in momentum state ${\bf
e}_i$ and only depends on the information at the site ${\bf x}$ at
time $t$.  The collision operation has the form\cite{fhp2}:
\begin{equation}
\Omega_i = \sum_{s, s'}(s' - s)\xi(s \rightarrow s')\prod_{j}
 N_{j}^{s_j}(1 - N_{j}^{s_j})^{1 - s_j},
\end{equation}
In the above equation, $s = (s_0,s_1, s_2, \cdot\cdot\cdot, s_6)$ and $s'$
represent the local states before and after collision and $\xi(s \rightarrow
s')$ is a Boolean matrix with expectation, $\langle \xi(s \rightarrow s')
\rangle$,   equal to the transition probability from state $s$ to $s'$, $ P(s
\rightarrow s')$. The conservation of mass
and momentum require the collision operator to satisfy
 $\sum_{i=0}^6
\Omega_i = 0$ and $\sum_{i=0}^6 {\bf e}_i \Omega_i = 0$. To allow the system
to approach a local equilibrium, the microscopic
collision transition probability $P$ usually
satisfies the   semi-detailed balance condition:
$\sum_{s}P(s \rightarrow s') = 1$

The simple collision processes, including the two-body head-on collision, three
body symmetric collision and one moving particle colliding with one rest
particle, are shown in Fig. 1.
If we let
 $f_i = \langle N_i \rangle$ be the ensemble-averaged particle distribution
($0\le f_i \le 1$),
one can prove that the local equilibrium, $f_i^{eq}$, can be described by the
Fermi-Dirac
statistics\cite{fhp2}:
\begin{equation}
 f_i^{eq} = \frac{1}{1+exp(\alpha + \beta {\bf e}_i \cdot {\bf v})},
\end{equation}
where $\alpha$ and $\beta$ are Lagrange multipliers
 determined by mass
and momentum conservation.

The local averaged  particle
density, $\rho$, the number density, $n$
 and the momentum, ${\bf j}$, are defined as follows:
\begin{eqnarray}
\rho({\bf x},t) \equiv m n({\bf x},t)\equiv
\sum_i f_i({\bf x},t) \nonumber \\
 {\bf j}({\bf x},t) \equiv \rho({\bf x},t) {\bf v}({\bf x},t)
\equiv \sum_i f_i({\bf x},t){\bf e}_i,
\end{eqnarray}
where   $m$ is the mass of one particle (usually equals to 1),
\[ f_i({\bf x},t) = \langle N_i({\bf x}, t) \rangle, \]
and ``$\langle \; \; \rangle$'' denotes an ensemble average.

In order to obtain the macroscopic hydrodynamic equation, one
assumes that $\epsilon = \frac{1}{L} \sim \frac{1}{T} \ll 1$, where
 $T$ is the macroscopic
characteristic time, $L$ is the macroscopic
characteristic length and $\epsilon$ is a small parameter. By taking
an ensemble
average of (1) and assuming  molecular chaos,
a Taylor expansion  gives the
continuum version of the kinetic equation
\begin{equation}
{\partial f_i \over \partial t} + {\bf e}_i \cdot \nabla f_i = \Omega_i,
\end{equation}
where $\Omega$ is the collision operator obtained by
 replacing $N_i$
by $f_i$ in (2). Using a Chapman-Enskog expansion (we will show
more details in the next section),
 it can be shown that
the system approximates the following
 fluid equations\cite{fhp1,fhp2}:
\begin{eqnarray}
{\partial n \over \partial t}
 &+& \nabla \cdot (n {\bf v}) = 0,  \nonumber \\
{\partial (n {\bf v}) \over \partial t}
&+& \nabla \cdot [n g(n ){\bf v}
{\bf v}]
= - \nabla p + \nu \nabla^2 (n {\bf v}),
\nonumber \\
p &=& \frac {1} {2} [n - g(n ){\bf v}^2].
\end{eqnarray}
Note that $g(n)$ is a function of density. $g(n)$
should be unity
for the correct
Navier-Stokes equation. This causes non-Galilean advection for velocity
field.  The
incompressible Navier-Stokes equations are recovered only in the low
Mach number limit when time, pressure and viscosity are rescaled by
the factor, $g$.  Because $g(n)$ depends on density, this rescaling is
consistent only for problems which have nearly constant density.
Another nonphysical term appears in the equation of state of the lattice gas
where the pressure depends on  the local velocity. This will cause some
nonphysical compressibility\cite{mong2,schen7}. Thus only small velocity flow
can be
approximated.

In Fig. 2, we present five contour plots for
vorticity at various times in the evolution of a Kelvin-Helmholtz instability
using a lattice
gas simulation\cite{loh}.  The simulation was
 on  $8192 \times 8192$ lattice
sites and  the initial Reynolds number was about 20,000. Since the
lattice gas
 method includes intrinsic noise, there is no need to introduce macroscopic
fluctuations to initiate
instabilities as is done
normally for
 finite-difference schemes. We observe that the initial vortex sheet naturally
separates into discrete vortices.
Gradually, vortices with the same sign
merge and ultimately form two large, oppositely rotating
vortices. During the merging process, one observes typical
spiral vortex patterns.
The viscous term in
the Navier-Stokes equation dissipates the vorticity.
Usually, viscosities in lattice gas models are not very small, and
vorticity dissipates rapidly. Because our simulation has a high
Reynolds number, the convection term still dominates. The vortex
dynamics consists mainly of vortex mixing and convection. After a
long time, the velocity field is accurately described as a
collection of discrete vortices.
The phenomena shown here are qualitatively
similar to other numerical simulations\cite{rog}. It is difficult
to quantify agreement between any pair of simulations for this
highly unstable flow.  However,
the simulation does demonstrate that the lattice gas method can
qualitatively generate meaningful results similar to other numerical methods.

\subsection{Lattice Boltzmann Method and the
Single Time Relaxation Approximation (STRA)}

In the lattice Boltzmann method, space and time are discrete as they are in the
lattice automaton method.
But instead of using  a bit representation for particles,
real numbers  represent the local
ensemble-averaged particle distribution functions and only kinetic
equations for the distribution functions are solved. This
contrasts with lattice gas models where molecules are represented
by bits and the motions of  individual particles are followed.
The lattice Boltzmann method ignores particle-particle
 correlations and often uses  simplified
collision operators.
However, even under these
simplifications they provide the correct evolution
 of the macroscopic quantities.
The lattice
Boltzmann method can be  viewed to some extent as a finite-difference technique
for solving the
kinetic equation. The Navier-Stokes equation can be recovered
in the long wavelength and low frequency limit. The
Boltzmann scheme is intermediate
 between the lattice gas method and the usual finite
difference algorithms.

As discussed in the last section, the lattice gas method
suffers from several major drawbacks, including a restriction to low Reynolds
numbers, a high noise-to-signal ratio, non-Galilean invariance and the velocity
dependence of the equation of state.
One crucial advantage of
the continuum description of the lattice Boltzmann method is
 that it eliminates most of the noise of the system
compared with the lattice gas method.  In addition, the lattice
Boltzmann method has
considerable flexibility in the choice of the local equilibrium
particle distribution.  In contrast, the Fermi-Dirac equilibrium is
the only distribution usually considered for the lattice gas method.
This additional freedom allows us
 to achieve desired physical properties, such as Galilean
invariant convection and a velocity independent equation of state.
The advantage of the lattice Boltzmann method over other finite difference
techniques for obtaining solutions of partial differential equations is that
they are  fully parallel, resulting in very fast codes. Given the increasing
availability of
parallel machines, there is a trend  to produce numerical codes that
exploit the intrinsic features of parallelism. The lattice Boltzmann
methods fulfill these requirements in a very straightforward manner.

The lattice Boltzmann model was first
proposed by McNamara and Zanetti\cite{macn-zan},
where they simply replace $N_i$ by $f_i$ in Eq. (1) without
changing the collision operator and streaming steps.
An important  refinement of the lattice Boltzmann method
was the recognition that
the ``exact''  collision integral is unnecessarily complex and numerically
inefficient if one aims to use it for solving macroscopic
equations.
Higuera and Jimenez\cite{hig-jim} proposed
to linearize the exact Boltzmann collision operator.
Expanding on this idea, two groups
nearly simultaneously offered the suggestion\cite{ccmm,qian,ccm}
that the exact collision operator can be, in effect, discarded,
provided that one adopts a collision operator that leads, in a
controllable fashion, to a desired local equilibrium state. By a
``desired'' equilibrium, we mean  one that (i)  depends only
upon the local fluid variables, which themselves can be computed
from the actual values of the local distribution at a point, (ii)
leads to the desired macroscopic equations (e.g., the Navier Stokes equation),
and (iii) admits desired additional properties,
such as
simplicity or removal of nonphysical lattice effects.

For developing a
magnetohydrodynamic (MHD) lattice Boltzmann
 method, Chen {\it et al.}
\cite{ccmm} offered the first suggestion that one could even simplify the
collision operator by
a single time relaxation, or STRA, process.
Subsequently, a similar method\cite{qian} was described, and referred
to as a ``BGK'' collision integral, in reference to the
more elaborate collision treatment of Bhatnagar, Gross and Krook\cite{BGK}.
The essence of the suggestion for the lattice Boltzmann method is that
the collision term, $\Omega(f)$, be
replaced
by the well-known single time relaxation approximation,
$\Omega(f) = -\frac{f-f^{eq}}{\tau}$.
The appropriately chosen equilibrium distribution is denoted by
$f^{eq}$, which depends
upon the local fluid variables. A lattice relaxation time, $\tau$,
controls the rate of approach to this equilibrium.
Later, Qian {\it et al.}\cite{qian} and Chen {\it et al.}
\cite{ccm} described a STRA
 method for hydrodynamics that incorporates a rest particle state which will
give the exact Navier-Stokes equations.

As mentioned before, the Boltzmann equation
for the hexagonal lattice  is constructed by replacing $f_i$, the
single-particle distribution,
for $N_i$, the particle occupation in Eq. (1):
\begin{equation}
f_i({\bf x}+ {\bf e}_i,t+1)=f_i({\bf x},t)+\Omega_i (f({\bf x},t)),
(i = 0,1 \cdot\cdot\cdot,6),
\end{equation}
where $\Omega_i=\Omega_i(f({\bf x},t))$ is a local collision operator.
 In the previous
section, we saw that Fermi-Dirac statistics are chosen and $f_i$ is
restricted to be in a range bounded by zero and one. For the lattice
Boltzmann method, the particle distribution does not have an upper
bound in general. To be consistent with particle distribution, we
require that $f_i \geq 0$.

Performing a Taylor expansion in time and space and taking the
long-wave and low frequency limit, we obtain the continuum form of the
kinetic equation  up to the second order from (7):
\begin{equation}
{\partial f_i \over \partial t} + {\bf e}_i \cdot \nabla f_i
+ \frac{1}{2} {\bf e}_i {\bf e}_i : \nabla \nabla f_i +
{\bf e}_i \cdot \nabla {\partial f_i \over \partial t} +
\frac{1}{2}{\partial^2 f_i \over \partial t^2}
 = \Omega_i
\end{equation}
Now we adopt the following multi-scaling  expansion\cite{fhp1}. The time  and
space derivative
is expanded as:
\[ {\partial \over \partial t} = \epsilon {\partial \over \partial t_1} +
\epsilon^2 {\partial \over \partial t_2}, \]
\[ {\partial \over \partial x} = \epsilon {\partial \over \partial x_1}, \]
where $\epsilon$ is the expansion paramter as defined in the last section.
The above formula implies that the diffusion time scale, $t_2$, is much slower
than the convection time scale $t_1$. Likewise, the one-particle distribution
function, $f_i$, is expanded, about the local equilibrium distribution
function,
\begin{equation}
f_i = f_i^{eq} + \epsilon f_i^{(1)} + \epsilon^2 f_i^{(2)}.\qquad\qquad
\end{equation}
Inserting $f_i$ into the collision operator,
we have,
\begin{equation}
\Omega_i (f)=\Omega_i(f^{eq}) + \epsilon {\partial\Omega_i(f^{eq})\over
 \partial f_j}f_j^{(1)}+
O(\epsilon^2).
\end{equation}
The Chapman-Enskog theory requires $\Omega_i(f^{eq}) = 0$.  Neglecting
higher order terms, we have the linearized form of the collision
operator:
\[ \Omega_i (f) = M_{ij}f_j^{(1)}. \]
Here, $M_{ij} = {\partial\Omega_i(f^{eq})\over \partial f_j}$.  If we
further assume that the local particle distribution relaxes to an
equilibrium state at a single rate: ${\partial\Omega_i\over\partial
f_j}= {-1\over \tau}\delta_{ij}$ with a universal  time scale $\tau$
\cite{ccmm}, we arrive at the following linearized form:
\begin{equation}
\Omega_i={-1\over \tau}(f_i-f_i^{eq}).
\end{equation}
Note that we have both $\sum_i\Omega_i=0$ and $\sum_i{\bf
e}_i\Omega_i=0$.  In order for the fluid to have Galilean-invariant
convection and a pressure which does not depend upon velocity, the
following equilibrium distribution, $f_i^{eq}$, is assumed:
\begin{eqnarray}
f_i^{eq}&=&\frac{\rho(1-\alpha)}{6} +
{\rho \over 3}{\bf e}_i \cdot {\bf v} +
{2\rho\over 3}({\bf e}_i)_\alpha({\bf e}_i)_\beta v_\alpha v_\beta-
{\rho \over 6}{\bf v}^2\\
f_0^{eq}&=&\alpha \rho-{\rho}{\bf v}^2.
\end{eqnarray}
where  $\alpha$ is a free parameter related to the sound speed as
shown below.
{}From (8), we can obtain the following equations,
\begin{eqnarray}
{\partial f_i^{eq} \over \partial t_1} + {\bf e}_i \cdot {\nabla}_1 f_i^{eq}
 = - \frac{f_i^{(1)}}{\tau}, \nonumber \\
{\partial f_i^{(1)} \over \partial t_2} + (1 - \frac{2}{\tau})[{\partial
f_i^{(1)} \over \partial t_1} + {\bf e}_i \cdot {\nabla}_1 f_i^{(1)}]
= - \frac{f_i^{(2)}}{\tau}.
\end{eqnarray}

After a simple algebra and using the definition in (4), the momentum equation
can be written as:
\[ {\partial \rho{\bf v} \over \partial t}  + \nabla \cdot \Pi = 0, \]
where the momentum flux tensor, $\Pi$, has the form:
\[ \Pi_{\alpha\beta} = \sum_i ({ \bf e}_i)_{\alpha}({\bf e}_i)_{\beta}
 [f^{eq}_i +
 (1-\frac{1}{2\tau})f_i^{(1)}], \]
where $({ \bf e}_i)_{\alpha}$ is the $\alpha$ component of the
velocity vector,  ${ \bf e}_i$.
Inserting (12) and (13) into  $\Pi$ and using the first equation in (14),  one
obtains, to zero order
in the small parameter of the Chapman-Enskog expansion,
\begin{eqnarray}
{\bf \Pi}^{(0)}_{\alpha \beta} & =  & \sum_i ({ \bf e}_i)_{\alpha}({\bf
e}_i)_{\beta} f^{eq}_i\nonumber \\ & = & 3 n \frac{1-\alpha}{6} \delta_{\alpha
\beta}+n v_{\alpha}v_{\beta}
\end{eqnarray}
and
\begin{eqnarray}
{\bf \Pi}^{(1)}_{\alpha \beta} =
-\tau\{\frac{\partial}{\partial t}{{\bf \Pi}^{(0)}_{\alpha \beta}}+
\frac{\partial}{\partial x_{\gamma}}\sum_{i}
({\bf e}_i)_ {\alpha}({\bf e}_ i)_ \beta ({\bf  e}_ i)_ \gamma f_{i}^{(0)}\},
\end{eqnarray}
to first order.
The final form of the macroscopic equations becomes,
\begin{eqnarray}
{\partial \rho \over \partial t}+\frac{\partial (\rho v_{\beta})}
{\partial x_{\beta}} &=& 0, \nonumber \\
n\frac{\partial v_{\alpha}}{\partial t}+nv_{\beta}\frac{\partial
v_{\alpha}}{\partial x_{\beta}} &=&-\frac{\partial p}{\partial
x_{\alpha}}+\frac{\partial}
{\partial x_{\beta}}(\frac{\lambda}{n}(\frac{\partial nv_{\gamma}}
{\partial x_{\gamma}}+v_{\alpha}\frac{\partial n}{\partial
x_{\beta}}+v_{\beta}\frac{\partial n}{\partial x_{\alpha}}))dvips \nonumber \\
&& + \frac{\partial}
{\partial x_{\beta}}(\mu(\frac{\partial v_{\beta}}
{\partial x_{\alpha}}+\frac{\partial v_{\alpha}}
{\partial x_{\beta}})), \nonumber \\
p  & = & \frac{1-\alpha}{2}\rho.
\end{eqnarray}
In the above equations,
 $p$ is the pressure and the sound speed, $c_s$,  is
 $\sqrt{(1-\alpha) \over 2}$.
The shear viscosity, $\mu$, is
 ${(2\tau-1) \over 8}$ and the bulk viscosity,
$\lambda$, is $\frac{(\tau-1/2)(2\alpha-1)\rho}{4}$.

Note that the above equation converges to the exact incompressible
Navier-Stokes equations only when the derivatives of the density in the second
viscosity term on the right hand side of the equation are small.
Since the gradients of the density are $O(u^{2})$ (see references
\cite{daniel1} and \cite{majda}), the unphysical terms in equation (16) are
$O(u^{3})$. Thus, although the physics contains compressibility effects, one
may come arbitrarily close to solving incompressible flows by reducing the Mach
number.

To solve incompressible fluid flows by traditional numerical methods,
such as finite-difference or finite-element, one must deal
with a Poisson equation for the pressure term that is induced by the continuum
condition and the momentum equation. Here we can see that the solution of the
Navier-Stokes equations can be obtained through the lattice Boltzmann equation
in (7) and the pressure effects on the momentum equation are controlled by
an equation of state.  Solution of the Poisson equation
 is actually avoided by relaxing the incompressibility requirement.
It can be argued
that lattice  methods are most closely related to  pseudo-compressible
algorithms\cite{pseudo}
for solving incompressible fluid flows.

There may still be some differences between the incompressible Navier-Stokes
equations and the macroscopic behavior of the discrete-velocity Boltzmann
equations because of the asymptotic nature of the Chapman-Enskog method. These
 differences could be attributed to  higher order
terms, such as the Burnett
terms,
 or as small deviations from the above relation for the kinematic viscosity.
 However,  Burnett
 terms are expected to become negligible as the global Knudsen number,
$Kn = \frac{l}{L}$,  becomes small, where $l$ is the mean free path.  Since the
Knudsen number is proportional to the Mach number divided by the Reynolds
number, the Burnett terms may be grouped
 with other ``compressibility"
 effects and should become small as the Mach number approaches zero for a fixed
Reynolds number.

It should be mentioned that the numerical implementation for (7)
is straightforward. At each time step, one calculates a new $f'_i$ on each
lattice:  $f'_i = f_i - \frac{f_i - f_i^{eq}}{\tau}$, where $f_i$ and
$f_i^{eq}$ are
all known functions. Then, $f'_i$ is advected to a new position, becoming a new
$f_i$. Using the new $f_i$, the density and velocity can be calculated again.

The lattice Boltzmann method can be tested by comparing with
pseudospectral results.
 In Fig. 3, we present
velocity contours for the
three-dimensional Taylor-Green vortex
(a cross section at $z = 64$) from
both numerical schemes  at times $t = 4$
and $t =10$. A
 body-centered-cubic lattice was used for
this calculation and the  system size was
$128^3$. The initial condition of the simulation and the comparison for other
quantities are given in detail in Ref.  \cite{sycjsp}.
We can see that the lattice Boltzmann method accurately reproduces the fluid
patterns. Note that for this flow at $t = 4$ the nonlinear interaction
dominates  and the vortex structure is being stretched. Whereas
at $t = 10$, the main feature is dissipation.
We see that the vortex structures
generated by the lattice Boltzmann method agree well with the spectral
method results both at early and later time, showing that the lattice Boltzmann
method can simulate both stretching and dissipation dynamics.

In Fig. 4, we show a snapshot during a two-dimensional simulation of flow past
a plate  done by L. Luo\cite{luo1}
using the lattice Boltzmann method. The attached vortices and vortex shedding
are evident, showing complicated flow patterns.

\section{Lattice Methods for Reaction-Diffusion Systems}

The lattice gas and the lattice Boltzmann methods have been used to simulate
phenomena whose macroscopic evolution is described by
reaction-diffusion equations.
Reaction-diffusion equations describe the evolution of the densities
of a set of species that interact through chemical reactions and
diffuse throughout an extended spatial domain.
In the case of isotropic diffusion,
the reaction-diffusion equations can be  written
as:
 \begin{equation}
{{\partial \rho_s}\over{\partial t}}-D_s \nabla ^2 \rho_s = R_s
,\qquad 1\le s \le M
\label{eq:RD1}
\end{equation}
where  $\rho_s({\bf x}, t)$ is the mass density of the species $s$
at time $t$ and position ${\bf x}$, $M$ is the number of species,
$D_s$ is the diffusion coefficient (which  we
assume   independent of $x$),
and $R_s$ is the
reaction term which  depends  on $\rho_s$ and
the densities of the other species that chemically react with $s$.
Usually $R_s$ is nonlinear in some of the densities, and couples
the different reaction-diffusion equations.

Experimental, numerical and analytical results \cite{refcas,reffife}
have shown that reaction-diffusion equations exhibit a
rich variety of behaviors, including, among others,
bistability, oscillations, chemical
waves, different types of
non-homogeneous  stationary states and
temporal and spatial-temporal chaos.
Also, reaction-diffusion equations have been used in
biological applications as models of self-organization and
pattern formation\cite{refmur}, and were originally proposed as a model
of morphogenesis\cite{reftur}. A striking feature
is that even very simple models are able
to produce amazingly complicated and interesting behaviors. Among
them,   the ``self-replicating'' spots recently observed in
numerical simulations \cite{refpear}, whose evolution resembles the process
of cell division.
Many of these phenomena have been reproduced by means of the lattice gas
\cite{refkap2,refkap1,refbrosl,refchop2,refchop1,physrep,gruner,chopard,refwu}
and the lattice Boltzmann methods \cite{refsil}.

Lattice gas simulations, due to their intrinsic noise,
have been used to probe the robustness of some of these
processes against the presence of internal fluctuations.
The ability to study the effects of fluctuations is especially
important near bifurcations, during the early stages
of the growth of structures or when the dynamics of the spatially
averaged macroscopic variables exhibits chaotic behavior.
The effect of fluctuations on the formation
and stability of Turing patterns was studied in Ref. \cite{refbrosl}
for a problem relevant to the biochemistry of the cell.
In Ref. \cite{gruner},
the structure and robustness against internal fluctuations
of a propagating chemical wave front
  was studied for  the Schl\"ogl model.
Direct simulations were also done to determine
the relative stability of the coexisting stable phases.
The observed phenomenology
agreed with the predictions of both theory and
experiments of optical bistable devices. The
domain formation and growth
from the unstable state were also examined, showing
that
the nonequilibrium correlation function and the corresponding dynamical
structure factor conformed to the predictions of scaling theory.
These results imply
 that the
lattice gas method gives the diffusive dynamics of the curved
interfaces  correctly,
and that the space-time fluctuations
produced by the automaton correspond to the fluctuations of a physical
system.\cite{gruner} The critical exponents characterizing the  properties
of reaction fronts
 as well as the role of fluctuations on the
 front formation
were also investigated   in Ref. \cite{chopard}. The structure of the front and
its velocity of propagation were also studied in Ref. \cite{lemarchand}.
The interaction between internal noise and deterministic
chaos has been recently analyzed by means of  lattice gas
simulations of reaction diffusion equations
with chemical kinetics given by the Willamowski-R\"ossler
model\cite{refwu}.
The interplay
among spatial degrees of freedom, system size and internal fluctuations
was also  studied in this work.

Lattice gas models were also used to study the effect
of particle correlations in reaction-diffusion systems.
In particular, the simulations carried out in  \cite{refkap2}
for the Schl\"ogl model showed that
when the frequency of reactive collisions  increases,  the dynamics
deviates from the phenomenological prediction
described by the Boltzmann
approximation.
These non-Boltzmann effects  were interpreted in terms of reactive
recorrelations. The
simulations  also showed the phase separation and wave
propagation phenomena expected for this system.
Static correlations have also been investigated for
another reactive lattice gas automaton yielding results in accordance with the
predictions of  a fluctuating-hydrodynamics model.\cite{weimar}

All these qualitative and quantitative
results support the use of the reactive lattice gas automaton
for investigating
  the reactive dynamics of spatially-distributed
systems at the mesoscopic level.

%A reactive lattice gas automaton model was also used to study adsorption
%processes in \cite{rikvold,wu1}. For example in \cite{wu1}
%a lattice gas  model for CO oxidation on metal surfaces
%was developed.  The model treats the dynamics of the adsorbed CO and O
%species explicitly and incorporates the effects of surface phase
%transformations through simple cellular automaton rules.  Nucleation and
%growth processes were observed and studied for this system.  Oscillations
%%%associated with
%surface spatial structure were also obtained when surface phase
%transformations were possible.

Both the lattice gas and the lattice Boltzmann methods
  can easily be extended to three spatial dimensions.
%,
%as we show in this paper for the latter.
The presence of a moving solvent can easily be
incorporated into the lattice Boltzmann scheme, as we have shown in
Ref. \cite{refsil}.
In particular, we have
 simulated a situation\cite{refsil} in which all species have the same
mean velocity as the solvent, and the evolution of this velocity
satisfies the Navier-Stokes equation, as described by the set:
 \begin{equation}
{{\partial \rho_s}\over{\partial t}}+ {\bf  \nabla \cdot} (\rho_s
{\bf v})
-D_s \nabla ^2 \rho_s = R_s
,\, \, \qquad 1\le s\le M
\label{eq:Ev1}
\end{equation}
 \begin{equation}
{{\partial \rho_0}\over{\partial t}}+ {\bf  \nabla \cdot} (\rho_0
{\bf v}) = 0,
\label{eq:Ev2}
\end{equation}
 \begin{equation}
\rho_0\left({{\partial {\bf v}}\over{\partial t}}+ {\bf v \cdot
 \nabla} {\bf v}\right)=- {\bf  \nabla \cdot}{\Pi_0}
+\rho_0\nu\nabla ^2{\bf v} +\rho_0\lambda {\bf  \nabla }(  {\bf\nabla \cdot
u}),
\label{eq:Ev3}
\end{equation}
where $\rho_0$ is the density of the solvent,
$\bf u$ is its mean velocity, $\Pi_0$ is the pressure
tensor, $\nu$ is the kinematic viscosity  and $\lambda$
is the bulk viscosity.
This set of equations allows
the study of pattern formation processes in the presence of
different types of flows.
Notice also that Eqs. (\ref{eq:RD1})
 reduce to Eqs. (\ref{eq:Ev1})  if ${\bf v}=0$.
The construction of a lattice gas cellular automaton capable of
reproducing the macroscopic behavior described by these equations
is not straightforward. We have also
developed a lattice Boltzmann scheme
 that simulates reaction-diffusion equations
in which each species is advected with a  different
but uniform and constant velocity in the presence of anisotropy
\cite{silana}
to study the interaction
between the Turing and the differential-flow-induced instabilities.
 \cite{rovmen1}

\subsection{Basic Features of Reactive Lattice Gas Models}

We  discuss here only the general
principles of multi-species
reactive lattice gas automata. For a more general description, see
Refs.
\cite{refkap2,refkap1,physrep}.
We consider spatially extended systems
composed of a number of species $X_{s}$, $s=1, \cdots, M$, which may
undergo chemical reactions of the type
\begin{equation}
\alpha_{1} X_{1}+\cdots+\alpha_{M} X_{M}{\rightleftharpoons}\beta_{1} X_{1}
+\cdots+\beta_{M} X_M\,.
\label{eq:R1}
\end{equation}
The aim is to construct a simplified molecular dynamics
which incorporates  the main characteristics of the real
collision and transport processes of the reacting systems,
yet is simple enough to perform simulations on large systems
composed of millions to billions of molecules.
The molecules of the reactive species reside at the nodes of the lattice
and to each molecule is associated a discrete velocity  vector.
In order to have a
computationally tractable memory requirement, an exclusion principle is assumed
 where
only one molecule of a given
species $s$ may reside at a node with a given velocity.
Physically, the exclusion principle prevents particle density
from building up at a local region of space, however it leads to some
restrictions on the range of processes that can be simulated with the
automaton.
Lattice gas automata for reaction-diffusion systems
without the exclusion principle have also been
considered \cite{refchop1},
but these models will  not be discussed  in this paper.

The real collision processes of the
reacting system are modeled by collision events
at the nodes of the lattice, $\br$, that are embodied in collision rules.
Both  elastic and reactive collision events must be incorporated into
these rules.
Since  in the dynamics of pure
reaction-diffusion systems
there are no fluid flows involved,  momentum
conservation is not required.
Most reactive lattice gas
models developed so far are for reaction-diffusion
systems in which the reactive species are
dispersed in some background medium, solvent or gel,
whose
dynamics is not followed by the model. Therefore,
the conservation of mass during collisions depends on whether or not
 the chemical species react
with the background species.
Assuming  that the concentration
of the solute species whose dynamics is followed is low   in comparison
with the background
medium, the overwhelming majority of non-reactive collisions will occur
between the solute species and the molecules
of this background.  Hence, it is
reasonable to assume
 that the elastic collisions occur independently for
each chemical species. In the lattice gas  automaton,
the diffusive dynamics   arises as a net effect of the
elastic collisions and the propagation between them, while
the chemical transformations are determined by the reactive collisions.

The elastic collisions of each species can be simulated by
  random rotations of the
velocity configuration or by random assignments of the velocities.
In order to preserve isotropy, some conditions must be  imposed on the
rotation probabilities\cite{refchop2}.
The random rotations will keep
 the exclusion rule automatically while
 some care must be taken in the case of
total randomization of the velocities.
Though the latter option  will
produce a better ``mixing'', it has been
shown that
the random rotations are  actually sufficient to simulate the
effect of diffusion\cite{refchop2}.

The reactive collisions  in the automaton
depend on the numbers of molecules of the various species
residing at the same node. They
are modeled by combining two operations:  local
reactions (\ref{eq:R1})
 that occur with preassigned probabilities which are
independent of the velocities, and a randomization of the
velocities.
 The net effect is
a reactive event where both particle numbers and velocities are changed,
just as in real reactive collisions.

Finally,
the molecules propagate  between collisions,
moving from one lattice node to the next.
Since elastic collisions and propagation processes are carried
out independently for each species,  they lead to independent
non-reactive dynamics
for different species.\cite{refchop2,refchop1,refkap1,physrep}
In order to obtain different  diffusion coefficients
for different species, it is necessary to repeat the
propagation and rotation steps a different number of times for each of them.
 For example,
in a system consisting of two reacting species,  $X_1$ and $X_2$,
applying $N_1$ propagation-rotation steps to the species
$1$ for every $N_2$ propagation-rotation steps of the species $2$
will give a ratio of diffusion coefficients
${{D_1}\over{D_2}}={{N_1}\over{N_2}}$.
The ability to  vary the diffusion coefficients of the
different species is an important part of the automaton
 in the study of Turing bifurcations where the
diffusion coefficient ratio is an essential parameter.

To compare the reactive lattice gas simulation results with
the solutions of reaction-diffusion equations,
an ensemble average must be taken. In actual numerical simulations,
the average over
realizations is replaced by an average in time or a spatial average over
neighboring lattice nodes (coarse-graining).
The  scheme  allows one to select
 the  reactive  lattice gas parameters so that the
dynamics of the averaged quantities is governed
by the desired macroscopic reaction-diffusion equations.
Additionally,  the  automaton  dynamics has
correlations  and  fluctuations that go beyond mean-field
descriptions. As we mentioned before,
this allows the study of many interesting
phenomena at the mesoscopic level.
Since  reactive collisions in the automaton determine the local
changes in the molecule numbers,   they control the local
number fluctuations. Hence the choice of the probabilistic rules
which govern these reactive collisions will affect the correspondence
between the fluctuations in the automaton and in the real
system that is being modeled.

\subsection{\bf Basic Features of the Reactive Lattice Boltzmann Method}

 Lattice
Boltzmann schemes for simulating different diffusion or reaction-diffusion
systems were introduced in Refs. \cite{refsuc}, \cite{refking}
 and \cite{refsil}.
In particular, in Ref.  \cite{refsil}
we  simulate the set
of Eqs. (\ref{eq:Ev1})-(\ref{eq:Ev3}) both in the presence and in the
absence of a moving solvent.
Here we will concentrate on the implementation of a lattice Boltzmann
scheme for this more general situation. This scheme
is basically the same as the one described for hydrodynamic problems,
though several lattice Boltzmann equations must be solved
simultaneously, one for the distribution particle of each reacting
species, $f^s_i$, $1\leq s\leq M$, and one for that of the solvent, $s=0$.
The main difference with respect to purely hydrodynamic
problems is reflected in  the collision operator,
which we write as
the sum of a reactive term, ($\Omega^{s,R}$) (which is zero for $s=0$)
and a non-reactive term, ($\Omega^{s,NR}$).
As  before, for $\Omega^{s,NR}$,
 we use the single relaxation
time approximation:
\begin{equation}
\Omega_i^{s,NR}({\bf x},t)=-{{f^s_i({\bf x},t)-f^{s,eq}
({\bf x},i,t)}\over{\tau_s}},\qquad 0\leq s\leq M,\label{eq:omnr}
\end{equation}
with, in principle, a different relaxation time, $\tau_s$, for each species,
with { equilibrium distribution functions}, $f^{s,eq}$, that can
 depend on
$\bf x$ and $t$ through the local densities,
\begin{equation}
n_s({\bf x},t)\equiv{{\rho_s} \over{ m_s }}\equiv
\sum_i f^s_i({\bf x},t),\qquad
0\leq s\leq M,
\label{eq:DENS1}
\end{equation}
where  $m_s$ is the unit
mass of species $s$,
 and through the
dimensionless mean velocity of the solvent
\begin{equation}
{\bf v}({\bf x},t)
\equiv {{1}\over{  n_0({\bf x},t)}}
 \sum_i {\bf e}_i f^0_i({\bf x}, t).
\label{eq:u01}
\end{equation}
 In particular, we choose the equilibrium distribution functions to satisfy the
following
conditions:
\begin{equation}
\sum_i  f^s_i({\bf x},t) = \sum_i
f^{s,eq}_i({\bf x},t) = n_s({\bf x},t),\qquad 0\leq s\leq M,
\label{eq:DENS}
\end{equation}
\begin{equation}
\sum_i {\bf e}_i  f^0_i({\bf x}, t) = \sum_i  {\bf e}_i
f^{0,eq}_i({\bf x},t) = n_0({\bf x},t){\bf v}({\bf x},t),
\label{eq:VEL0}
\end{equation}
\begin{equation}
 \sum_i {\bf e}_i f^{s,eq}_i({\bf x}, t) =  n_s {\bf v}({\bf x},t),
\qquad 1\leq s\leq M.
\label{eq:COMP}
\end{equation}
Eq. (\ref{eq:DENS})  represents the conservation of  mass
 for a non-reactive system, while Eq. (\ref{eq:VEL0}) represents the
conservation
of momentum for the solvent. Momentum  is not conserved for
each reacting species, neither do we require the conservation of momentum
for all the species as a whole.
Each species is
advected by the solvent flow and for this reason we impose Eq. (\ref{eq:COMP})
on the corresponding equilibrium distributions.
In order to satisfy Eq. (\ref{eq:COMP}), we choose
 equilibrium distributions  that are related by:
\begin{equation}
f^{s,eq}_i({\bf x},t)=
{{n_s}\over{n_0}} f_i^{0,eq}({\bf x},t).
\label{eq:EQ22}
\end{equation}
To obtain  the correct Navier-Stokes equation for ${\bf v}$,
we use the same equilibrium distribution for the solvent as the one
defined by Eqs. (12)-(13) with $\alpha=1/7$.
%:
%\begin{equation}
%{{f_0^{eq}({\bf x},i,t)}\over{n_0({\bf x},t)}}=\cases{
%{{1-\alpha}\over {7}} +{{1}\over{3}}{\bf e_i . \tilde u}
%({\bf x},t)+
%{{2}\over{3}} {\bf e_i e_i ::  \tilde u}
%({\bf x},t) {\bf \tilde u}({\bf x},t)
% - {{ {\bf\tilde u}^2}\over{6}},&for $1\leq i\leq 6$;\cr
%{{1+6\alpha}\over {7}}- {{\bf\tilde u}^2},
%&for $i= 0$;\cr}
%\label{eq:EQ2D}
%\end{equation}
%in the two-dimensional case, and
%\begin{equation}
%{{f_i^{0,eq}({\bf x},t)}\over{n_0({\bf x},t)}}=\cases{
%{{1-\alpha}\over {7}} +{1\over 3}{\bf e_i .  u}
%({\bf x},t)+
%{1\over 2} {\bf e_i e_i ::   u}
%({\bf x},t) {\bf  u}({\bf x},t)
% - {{ {\bf v}^2}\over{6}},&for $1\leq i\leq 6$;\cr
%{{1-\alpha}\over {56}} +{{1}\over{24}}{\bf e_i .  u}
%({\bf x},t)+
%{{1}\over{16}} {\bf e_i e_i ::   u}
%({\bf x},t) {\bf v}({\bf x},t)
% - {{ {\bf v}^2}\over{48}},&for $7\leq i\leq 14$;\cr
%{\alpha}- {{{\bf v}^2}\over{3}}
%,&for $i= 0$;\cr},\label{eq:EQ3D}
%\end{equation}
%in the three-dimensional case, where we
%consider a cubic lattice of unit size, six velocity vectors
%of norm one, ${\bf e}_i$, $1\leq i\leq 6$, connecting nearest neighbors,
% eight velocity vectors of norm $\sqrt 3$, ${\bf e}_i$, $7\leq i\leq 14$,
%connecting second-nearest neighbors, and
% a zero velocity vector, ${\bf e}_0 = 0$. In Eqs. (11),(12)
%and (\ref{eq:EQ3D}) $\alpha<1$ is
%a free parameter that we choose equal to zero for the
%two-dimensional case and to ${1\over 8}$ for the
%three-dimensional case.
The equilibrium distributions
 of the other species can be obtained afterwards using Eq. (\ref{eq:EQ22}).

For  $\Omega_s^{R}$ we choose the
 simple and isotropic form:
 \begin{equation}
\Omega_s^{R}={{\tilde R_s}\over{7}}, \label{eq:omr}
\end{equation}
where ${{\tilde R_s}}=
{{R_s}\over{m_s}}$, with $R_s$ the same as in Eq.
(\ref{eq:Ev1}).
 This choice is independent of the particle speed, ${\bf e}_i$, and is thus a
 simplification of what occurs at the microscopic
level. For a discussion of this, see Ref. \cite{refsil}.

In order to show how  the results
obtained by solving the coupled lattice Boltzmann equations
are related to the
solutions of Eqs. (\ref{eq:Ev1})-(\ref{eq:Ev3}), we need
to derive the evolution
equations for  the densities, $n_s$, and the velocity, $\bf u$,
as explained in previous sections.
In this case, in order
to recover the correct macroscopic equations, we need to
assume the multi-scaling
${{\partial}\over{\partial t}}\sim
\epsilon^2$,  ${{\partial}\over{\partial x}}\sim
\epsilon$, $\tilde R_s \sim
\epsilon^2$, $\vert{{\partial n_0}\over
{\partial {\bf x}}}\vert\ll \vert{\bf v}
\vert\sim \epsilon$, where $\epsilon$ is a small parameter.
 Once these expansions are introduced, summing
the lattice Boltzmann equations over all velocities,
we obtain the reaction-diffusion equations for $1\leq s\leq M$
and the continuity equation for the solvent, $s=0$, if we
keep terms  up to order $\epsilon ^2$.
 Multiplying the lattice Boltzmann equation
 for the solvent, $s=0$,  by ${\bf e}_i$ and summing
over the velocities, we obtain the Navier-Stokes equation
if we
keep terms  up to order $\epsilon ^3$.
Under these scaling conditions, the evolution
of the macroscopic quantities is governed by the following set
of equations. (See Ref. \cite{refsil}
for more details):
\begin{equation}
{{\partial n_s}\over{\partial t}}+{\bf\nabla}\cdot(n_s {\bf v})-
{{3(1-\alpha)}\over{7}} (\tau_s - {1\over 2}) \nabla ^2 n_s
= \tilde R_s
\label{eq:EVONS}
\end{equation}
\begin{equation}
{{\partial n_0}\over{\partial t}}+{\bf\nabla}\cdot(n_0 {\bf v})
= 0
\label{eq:EVON0}
\end{equation}
  \begin{equation}
n_0\left({{\partial {\bf v}}\over{\partial t}}+ {\bf v}\cdot
{\nabla} {\bf v}\right)=- {\nabla}\cdot
{\widetilde{\Pi}_0}+
{{n_0}\over{4}}(\tau_0-{{1}\over{2}})\nabla ^2{\bf v}
+{{n_0}\over{4}}\left(\tau_0\left
(2-{{12}\over{7}}(1-\alpha)\right)-1\right)
{\bf  \nabla  } ({\nabla}\cdot u),\label{eq:EVOV0}
\end{equation}
%\begin{equation}
%n_0\left({{\partial {\bf v}}\over{\partial t}}+ {\bf v}
%{\bf .\nabla} {\bf v}\right)=- {\bf  \nabla .}
%{\widetilde{\Pi}_0}+
%{{n_0}\over{3}}(\tau_0-{{1}\over{2}})\nabla ^2{\bf v}
%+{{n_0}\over{3}}\left(\tau_0\left
%(2-{{9}\over{7}}(1-\alpha)\right)-1\right)
%{\bf  \nabla  } ({\nabla .u}),\label{eq:EVOV03D}
%\end{equation}
where ${\widetilde{\Pi}_{0_{kj}}}=n_0 {{3(1-\alpha)}
\over{7}}\delta_{kj}$ is the  pressure tensor.
Eqs. (\ref{eq:EVONS})-(\ref{eq:EVOV0})
are of the same form as Eqs. (\ref{eq:Ev1})-(\ref{eq:Ev3}). From them we can
obtain
the transport coefficients in terms of the various $\tau _s$.
We obtain
$ D_s={{3(1-\alpha)}\over{7}} (\tau_s - {1\over 2})$ for $s\ne 0$,
$\nu={{1}\over{4}}(\tau_0-{{1}\over{2}})$ and
$\lambda={{1}\over{4}}\left(\tau_0\left
(2-{{12}\over{7}}(1-\alpha)\right)-1\right)$.
% in the two-dimensional case and $\nu=
%{{1}\over{3}}(\tau_0-{{1}\over{2}})$ and
%$\lambda={{1}\over{3}}\left(\tau_0\left
%(2-{{9}\over{7}}(1-\alpha)\right)-1\right)$
%in the three-dimensional one.

\subsection{Numerical Simulations}

 We show now the results of some simulations of reaction-diffusion systems
using the lattice gas and the lattice Boltzmann
techniques. All of them correspond to chemical kinetics described by
 the Sel'kov model \cite{refsel}:
\begin{equation}
A \bbrel{k_1}{\rightleftharpoons}{k_{-1}} X,
\hspace{0.1in} X + 2Y \bbrel{k_2}{\rightleftharpoons}{k_{-2}} 3Y,
\hspace{0.1in} Y \bbrel{k_3}{\rightleftharpoons}{k_{-3}} B.
\label{eq:selkov}
\end{equation}
This is a model of biological interest that
was originally derived from a study of  oscillations in
 the glycolysis.\cite{refsel}
As in previous papers (\cite{refkap1},\cite{refbrosl}, \cite{refsil}),
all simulations have been done assuming
 that the densities  of $A$ and $B$, $\rho_a$ and $\rho_b$,
are fixed by external flows and can be treated
 as parameters that do not depend on time or space.
In this way, only the dynamics of the intermediate
 species, $X$ and $Y$, is followed by the model.
The corresponding macroscopic equations for the densities,
$\rho_1$ of $X$ and $\rho_2$ of $Y$, are two reaction-diffusion
equations of the form (\ref{eq:RD1}) with dimensionless reaction terms
of the form:
$\tilde R_1 =
 k_1 n_a -  k_{-1}n_1 -  k_2 n_1 n_2^2 +  k_{-2}
n_2^3$ and
 $\tilde R_2 =  k_{-3} n_b -
 k_{3} n_2 +  k_2 n_1 n_2^2 -  k_{-2}
n_2^3$, where
the constants, $k_i$,
determine the reaction rates.
This simplified model still
has a complicated bifurcation structure and contains
steady states, oscillations and  bistabilities.\cite{refsel}

%  The first simulations that we describe here
%were carried out on a square lattice.
% and the $(25 \times 25)$
%reaction probability matrix $P(\alpha_X \alpha_Y|\beta_X \beta_Y)$
%that specifies the transition probability for the reaction,
%$\alpha_X X +\alpha_Y Y \rightarrow \beta_X X +\beta_Y Y$, was
%constructed to conform to the reaction mechanism.  In
%addition, the elements of the probability matrix satisfy conditions
%such that the mean field rate law derived from the automaton dynamics
%corresponds to  Eqs. (\ref{selkovlaw1}) and (\ref{selkovlaw2}).

One of the problems that have been
  investigated  by lattice gas  simulations of the Sel'kov model
was whether spontaneous fluctuations in
the system can produce nuclei that exceed a critical radius and thus
generate spatial structures.\cite{refkap1}
%The simulations were carried out for a selection of rate coefficients
%which correspond to   homogeneous
%kinetics within the excitable region.
The simulations showed that
threshold-exceeding local
perturbations
  produce chemical waves
of excitation that propagate through the reacting medium.
If two
such chemical waves meet, they annihilate.  If the medium is disturbed or
if it
is inhomogeneous, the rings of excitation may be broken and the free ends will
curl to form spiral waves.  These phenomena correspond to
 commonly observed chemical patterns\cite{refmur,field}.
We show in Fig. 5
the result of a lattice gas simulation carried out on a square lattice
in which this phenomenon can be observed.
  In
addition to an externally induced perturbation one can also observe the
formation and growth of spontaneous nuclei giving rise to a complex
dynamic spatial state that consists of annihilating wave fragments and
newly-formed nuclei and rings of excitation.

Another phenomenon that was studied  \cite{refkap1,refbrosl}
by means of lattice gas simulations of the Sel'kov model is
the so called Turing instability \cite{reftur}.
This is a
spatial symmetry-breaking instability which
is driven by diffusion and that gives rise to the formation
of stationary patterns.
%in the automaton was
%studied  in
% by choosing rate
%constants, $\rho_A$ and $\rho_B$ concentrations and diffusion coefficients
%to drive a Turing bifurcation in the Sel'kov reaction-diffusion equations.
%The parameters of the simulation are such that
%the reaction-diffusion equations possess a unique
%homogeneous and stationary
%solution which is linearly stable to small  inhomogeneous
%perturbations if the diffusion coefficients of $X$ and $Y$ are equal.
%If
%the ratio of diffusion coefficients
%exceeds a certain threshold, however,
%the homogeneous
%steady state becomes unstable and the system bifurcates to an
%inhomogeneous steady state.
We show in Fig. 6 the formation of these
{ Turing patterns} from random initial
conditions as reproduced by a lattice gas simulation.
 For the kinetic parameters of the simulation, the instability occurs if
the ratio of diffusion coefficients, $D_1/D_2$, is bigger than
$16.2$. The simulation
was done for $D_1=25\times D_2=12.5$
  $c^2$/time step, with $c$ the lattice separation.  The random initial
condition was chosen so that the average concentrations of the $X$
and $Y$ species were equal to the homogeneous fixed point values.
The system evolved from this random initial state to a  state
with hexagonal symmetry whose wavelength compares well with the value
predicted from the linear stability analysis.\cite{refbrosl}
The Turing pattern is
stable for long time periods but is subject to the effects of fluctuations,
as observed in the figure.
%Fig. 3 shows the time evolution of the spatially averaged
%concentrations of $X$ and $Y$ as a plot in the concentration plane.

We show now the results of some numerical simulations using the lattice
Boltzmann method.
We present in Fig. 7 a comparison of the formation of ``Turing'' patterns
in the absence (first row) and presence (second row) of a moving solvent,
for a system described by Eqs. (\ref{eq:Ev1})-(\ref{eq:Ev3})
with $M=2$ and reaction kinetics described by the Sel'kov model as
in the previous lattice gas simulations.
Both simulations depicted in Fig. 7 correspond
to the same rate constants, $k_i$, (which are
equal to the ones of Fig. 6), system size (64x64 grid points on
an hexagonal two-dimensional lattice),
boundary conditions (periodic) and diffusion coefficients
($ D_1=0.2914$ and
$ D_2=0.0170$). In the absence of flow (first row),
the system settles onto an hexagonal array of spots of high
$Y$ concentration on top of a lower concentration background,
in accordance with what is expected for the parameters
we used. When a
flow,  ${\bf v}({\bf x},t) =  v_x (y) \hat x$,
%with $v_x (y-{{N_y}\over 2}) = -v_x({{N_y}\over 2}-y)$,
is imposed,
 some spots of roughly the same wavelength as before arise ($t=4000$).
However, these spots
 move and deform in time, until they turn into a set of stripes, which is
the only pattern compatible with the breaking of
 symmetry that the flow introduces.
Meanwhile, the density difference between the maximum
and the minimum values decreases with time and finally remains
 constant. As mentioned before,  designing a
lattice gas automaton capable of modeling the problem with flow is not
straightforward.

It is evident from the figures that one of the
differences between lattice-gas and lattice-Boltzmann simulations
is the presence of noise in the former. To make this more evident,
we compare in Fig. 8 the results of the same simulation done
using both techniques.
We show in Fig. 8 different snapshots of the density of species
$Y$, ($n_2$), at different times, for a two-dimensional case with no
moving solvent.
The first row corresponds to a
simulation using the lattice gas method, while the second row corresponds
to a simulation using the lattice Boltzmann one. The reaction rates,
$ k_i$, the system size (256x256 grid points on an
hexagonal lattice), the boundary conditions
(periodic) and the ratio of diffusion coefficients ($D_1/D_2=8$)
are the same in both cases. The diffusion coefficients are very similar
 ($D_1=1$ and $D_2=0.125$ for the lattice gas; $D_1=1.0286$ and
$D_2 = 0.1286$ for the lattice Boltzmann scheme)
while the initial conditions and the color map
are slightly different.
In both cases we observe a spot of high $Y$ concentration in the
middle of a region of lower concentration. The spot is initially slightly
deformed and, after a while, it splits into two new spots of high
concentration in a sea of low $Y$ concentration.
While the separation is clear and neat in the lattice Boltzmann simulation,
the presence of noise is evident in the lattice gas simulation. However, if
we averaged this noise, we would be able to recover the same
kind of picture as in the Boltzmann simulation. The difference in the
corresponding wavelengths is due to the slight difference in the
diffusion coefficients of both simulations and to the
effect of fluctuations in the lattice gas.
The phenomenon depicted in this picture is the so called
``spot-replication'' that has been recently observed using a finite-difference
method
in Ref. \cite{refpear}.

\section{Chemical Reaction Process On a Surface}

Chemical reaction processes involving solids, aqueous fluids and
gases are commonly spatially distributed in both geologic and
engineered systems.  Intrinsically coupled flow and chemical
reaction processes are a common factor.  In geologic systems,
increasingly detailed analytical capabilities provide complex data
sets for the chemical and mineralogical distributions resulting from
such processes.  However, such observations are dominantly
concentrated at microscopic scales, while the goal of
understanding processes involves much larger scales.  Thus, a
simulation capability which integrates the detailed observations
and provides a link to macroscopic and regional scales for
mineral-fluid interaction as a function of time and space is critical to our
understanding of a wide variety of basic geochemical processes.
There are also a variety of practical applications which demand
sophisticated modeling capabilities, including applications to
geologic systems such as petroleum reservoirs, environmental
contaminant transport, high-level radioactive waste repository
evaluation and ore deposit formation.  In petroleum reservoirs,
applications of interest include coupling between oil migration,
secondary recovery processes, and the evolution and/or
manipulation of porosity.  Similarly, reactions coupled to ground
water flow and contaminant migration are important for
environmental concerns.

As chemical reaction models have become increasingly
comprehensive to include such complexities, it has been
recognized that new modeling approaches are required to
understand many pressing issues.  Interrelated factors such as
rock texture, mineral distribution, multiphase flow, and pore or
fracture network geometry all affect the simultaneously interrelated
processes of mineral dissolution, deposition, and mass transport.
In addition, reaction kinetics can significantly affect the integrated
results of such processes.  For bulk reactions or large scales of
tens to thousands of meters, models of geochemical processes
have provided significant insights using averaged or
phenomenological descriptions of the permeability, fluid flow fields,
mineral distributions, and fluid composition \cite{Bassett,Lichtner,Ague}
(and references
therein).  There are also models for reactions on the molecular
scale \cite{Lasaga1}.  Between these two scales,
however, there is a paucity of general  flexible models of fluid-
rock interaction.  This intermediate scale is particularly important to
a quantitative understanding of both geochemical and flow
processes in natural porous media because it is precisely the scale
at which most detailed analytical, experimental, and descriptive
methods provide information.  Lattice gas and lattice
Boltzmann hydrodynamic models which integrate chemical
transport and reaction processes are being developed for
application to such intermediate scales.

The simplest models for chemical processes in a porous media
involve diffusion in the solvent occupying channels or pores and
reactions at solid surfaces.  Initial applications of lattice models
have been to systems involving solutes in an aqueous solvent
\cite{wells2,janecky1}.  Thus, there
is a distinction made between ``particles'' in the lattice gas or
lattice Boltzmann sense
and the chemical perspective of units of solution containing both
solvent and solutes.  The models also assume homogeneous
solution transport, as is commonly assumed for geochemical
calculations \cite{Helgeson}.

In a geologic pore network, chemical reactions at surfaces may
take several forms.  Surface reactions may involve significant mass
transfer via dissolution and/or precipitation which modifies the
pore network structure and thus the hydrodynamic environment.
When the components of interest are at minor or trace
concentrations, sorption and desorption reactions become
important and the pore network may be either fixed or variable.
Catalytic reactions of solutes at solid surfaces could also be
modeled using this approach.

Solutes are transported down a concentration gradient by diffusion.
To simulate diffusion in a lattice gas model, a concentration of a solute
was assigned to each open space node \cite{wells2}.
Where the mass of solute or changes in the mass of solute are
small relative to the total mass of solution, only minor perturbations
of momentum for the lattice ``particles'' are involved which can be
ignored.  This approach to solution composition tracking is similar
to the energy tracking approach of
 Sero-Guillaume and Bernardin\cite{Sero}.  In lattice gas
 models, each ``particle'' is assigned a concentration
equal to that of the node at which it is residing.  During each time
step, ``particles'' travel to new nodes and the concentration of those
nodes is assigned the concentration of the arriving ``particle''.  If
more than one ``particle'' arrives at a node during a time step, the
resulting concentration at that node becomes the mean
concentration of all ``particles''.  During the next time step, all
``particles'' carry the mean concentration to their next destination.
This method of simulating diffusion contrasts with models for
 inter-diffusion of different types or species of ``particles'', primarily by
considering solvent flow separately from solute diffusion and
transport.  This distinction between the solvent flow in the form of
``particles'' and solutes as variable terms attached to the ``particles''
results in considerable flexibility for the models, including
multicomponent solute extensions.  This approach was examined
by performing a set of calculations of diffusion in a closed box in
which mean concentration profiles at different time steps for the
lattice gas calculation are in good agreement with  Fickian diffusion
profiles calculated using a finite difference technique\cite{wells2}.
This calculation also illustrated the level of noise
inherent in lattice gas calculations, in which even the average was not
completely smooth for a 64x64 node box.  The rate of diffusion was
also used to delineate the space-time relationships of models that
are being applied to real physical problems.  For instance, Wells
and others \cite{wells2} calculated a space to time ratio of 0.002 nodal
spacing in meters per second, using a geologically
reasonable diffusion coefficient of $10^{-9}
m^2/sec$.

In comparison to the lattice gas results, the lattice Boltzmann
 diffusion calculation
produces smooth profiles, with no variations along the diffusion
front \cite{janecky1}.  Extension of these models to
multicomponent-multiphase systems with independent diffusion
coefficients is relatively straightforward.  In fact, the present
lattice Boltzmann
model code includes the capability for up to 200 solute components and
binary solute reactions \cite{refsil}.

Surface reactions involving dissolution and precipitation are
simulated by allowing wall nodes to serve as sources or sinks for
mass of a dissolved component \cite{wells2}.
Whenever a ``particle'' collides with a wall, mass may be
exchanged, thus increasing or decreasing the local concentration
in solution depending upon the saturation state of the fluid.  Any
thermodynamic or kinetic reaction model can be integrated into the
lattice models.  For example, Wells and others\cite{wells2} utilized a
representation of a transition state rate law \cite{Helgeson}
 in an example calculation of surface reactions coupled with
flow through a simple porous network which included two mixing
channels and a dead-end pore (Figure 9).  The concentration field
that results from the reaction of a wall mineral with an initially
undersaturated fluid is clearly inhomogeneous.  The concentration
within the principle flow channel is relatively low, whereas the
concentration in dead-end pores and around obstructions is
relatively high. This type of simulation has important implications
for increasing our understanding of the relationship between
porosity and permeability in dynamic systems in which the porous
structure of a rock is altered by mineral dissolution and growth.
The effect of reaction and transport heterogeneities on the overall
evolution of fluid composition and rock porosity and permeability is
not well understood at this time, but is certainly a function of fluid
velocity, rock composition, and the geometry of the porous
network.  The lattice approaches have great potential to aid in
investigations of these phenomena.

Many mineral dissolution reactions intrinsically are spatially
heterogeneous.  For example, crystallographic defects are sites of
excess strain energy which constitute regions especially favorable
to dissolution \cite{Lasaga2} .  Such heterogeneous
surface reactions were also simulated with the lattice gas model of
Wells and others\cite{wells2} by defining zones of higher solubility.
This approach is a simple example of the broad range of systems
which can be modeled by distributing variable reactive parameters
across the lattice.  In a simple case, the lattice models can
effectively model transport-controlled dissolution in which a
significant rate-limiting step is the diffusion of the dissolved
component away from a mineral surface.  This is possible because
the model explicitly treats both the fluid and the mineral surface.

Two types of chemical reaction processes in flowing systems have
begun to be applied to evaluation of geochemical processes
through application of the approaches defined by Wells and others
\cite{wells2} for lattice gas automata models, and subsequently
enhanced for lattice Boltzman models \cite{janecky1}.
Sorption and desorption reactions are the focus of work with fixed
boundary conditions between solution and solid \cite{janecky2}.  Relatively
simple precipitation and dissolution
reaction processes have been abstracted\cite{Shanks}, where moving boundary
conditions and growth of
mineral/rock structures are being investigated.

Evaluation of how chemical sorption processes measured in the
laboratory relate to large scale transport in aquifers and the
unsaturated zone integrates a variety of processes.  In particular,
heterogeneous pore networks and distribution of sorbing minerals
result in both physical and chemical dispersion.  Models of these
coupled processes are necessary to examine competing
phenomena and to define simulation approaches for scaling
sorption processes into models of regional hydrologic systems.
Initial detailed models have been developed for sorption/desorption
chemical reactions at solid surfaces\cite{janecky2},
including multicomponent sorption/site competition as a function of
space and time in heterogeneous pore network structures, where
hydrodynamic transport, solute diffusion and mineral surface
processes are all treated explicitly.  The simplicity and flexibility of
the approach allow study of the interrelationships between fluid
flow and chemical reactions in porous materials at a resolution
that has not previously been computationally possible
(Figure 10).

In the case of growth of chimney structures at seafloor
hydrothermal vents, precipitation and dissolution based models are
being utilized \cite{Shanks}.  Initial chemical transport
and interaction processes have been developed for two-dimensional
models of simple jets (laminar to turbulent flows) and CaSO$_4$-
temperature components.  Development of anhydrite (CaSO$_4$)
chimneys is being modeled as a first step in understanding the
development hydrothermal precipitates during mixing between
hydrothermal fluid (up to $350^{o}$C) and seawater
(approximately $2^{o}$C) at the
seafloor interface.  Additional components and solids (especially
sulfide minerals) are being added as our experience in
implementing chemical reaction codes on massively parallel
computers develops.

Finally, a reactive lattice gas automaton model was also
used to study adsorption processes in \cite{rikvold,refwu1}.
For example in \cite{refwu1} a lattice gas model for CO
oxidation on metal surfaces was developed.
The model treats the dynamics of the adsorbed CO and O species
explicitly and incorporates the effects of surface
phase transformations. Nucleation and growth processes were
observed and studied for this system. Oscillations associated
with surface spatial structure were also obtained when
surface phase transformations were possible.
The developed in \cite{refwu1} model can serve as a basis for more
detailed studies of the reactive dynamics of such systems.

\section{Interfacial Tension and Multiphase Fluids in Porous Media}

The numerical simulation of two-phase flows is an interesting
and challenging
problem  because of the difficulty of modeling interface
dynamics and
the importance of potential
applications, including flow through porous media, viscous
fingering
and dendrite formation. Traditional
finite difference and finite element schemes have been used
for simple boundaries,  but are difficult to apply to
very complicated interfaces.
Recent developments in lattice
methods
 have provided the possibility of simulating
 complex two-phase phenomena.

Rothman and Keller\cite{roth2} were the first to extend the single-phase
FHP\cite{fhp1} lattice gas model
to simulate multi-phase fluid problems. They introduce
the colored particles to distinguish among phases.
A nearest-neighbor particle
interaction was used to implement interfacial dynamics, such as Laplace's
formula for surface tension.  Later, Somers and Rem\cite{somers},
and Chen {\em et al.} \cite{schen3} extended the original colored-particle
scheme by introducing colored holes.  It has been shown\cite{schen3}
that the colored-hole lattice gas method
extends the original nearest neighbor particle interactions
to several lattice lengths, leading to a Yukawa potential.  Moreover,
the colored-hole scheme carries purely local information in its particle
collision step, reducing the size of the look-up table in the algorithm
and consequently speeding up the simulation.

Although two-phase lattice gas algorithms are able to produce interesting
surface phenomena, they are difficult to compare quantitatively
with experiments and other numerical simulations due to noise
induced by particle fluctuations.
Combining the lattice Boltzmann ideology with their original idea of using
particle-particle
nearest neighbor interactions, Gunstensen {\em et al.} \cite{gun} first
proposed a lattice Boltzmann method for solving two-phase fluid flows. An
important contribution of this model is the explicit introduction of a
perturbation step (shown in detail below) so that
Laplace's formula at an interface can be approximated. Recently,
Grunau {\em et al.} \cite{daryl}  have extended this model by including density
and viscosity
variations. In the following sections, we briefly summarize this
new model and present some simulation results, including the validation of the
method and its application to multiphase fluid flows through porous media.

Denote $f_i({\bf x},t), f_i^{(r)}({\bf x},t)$ and $f_i^{(b)}({\bf x},t)$ as
the particle
distribution functions at space ${\bf x}$ and time $t$ for total, red,
and blue fluids, respectively.  Here $i = 0, 1,\cdot\cdot\cdot,N$,
where $N=6$ is the number of moving particle directions on a hexagonal
lattice, and $f_i = f_i^{(r)} + f_i^{(b)}$.  The lattice Boltzmann equation
 (7) can be extended to simulate
red and blue fluids:
\begin{equation}
f_i^{k}({\bf x} + {\bf e}_i , t +1) = f_i^{k}({\bf x}, t) +
\Omega_i^{k}({\bf x}, t), \label{eq:1}
\end{equation}
were $k$ denotes either the red or blue fluid, and $\Omega_i^{k} =
(\Omega_i^{k})^{1} + (\Omega_i^{k})^{2}$ is the collision operator.
The first term of the collision operator, $(\Omega_i^{k})^{1}$,
represents the process of
relaxation to local equilibrium as was done for
 the single phase in (7).  A linearized
collision operator with a single time relaxation parameter, $\tau_k$,\cite{ccm}
can be imposed for each phase,
\[ (\Omega_i^{k})^{1} = \frac{-1}{\tau_k}(f_i^{k}-f_i^{k(eq)}). \]
Here, $f_i^{k(eq)}$ is the local equilibrium
state depending on the local density and velocity.  $\tau_k$ is
the characteristic relaxation time for species $k$. The choice of $\tau_k$ will
allow one to have different viscosities for different phases.
The distribution function
must satisfy  conservation of mass and momentum:
\[ \rho_r = \sum_i f_i^{r} = \sum_i f_i^{r(eq)}, \]
\[ \rho_b = \sum_i f_i^{b} = \sum_i f_i^{b(eq)}, \]
and
\[ \rho {\bf v} = \sum_{i, k} f_i^{k} {\bf e}_i =
\sum_{i, k} f_i^{k(eq)} {\bf e}_i. \]
Here, $\rho_r\ {\rm and}\ \rho_b$ are densities of the red and blue fluids
respectively, $\rho =
\rho_r + \rho_b$ is the total density; and ${\bf v}$ is the local velocity.
The first
two equations represent the conservation of mass for each phase. The
third equation represents the conservation of total momentum.
Following the STRA model in Sec. 2 \cite{sycjsp},
the following equilibrium distribution for both red and blue fluids can be
used,
\begin{eqnarray*}
%f_i^{r(eq)}&=&\rho_r\left [\frac{1}{6+m_r}+\frac{1}{3}({\bf e}_i)_\alpha
%v_\alpha+\frac{2}{3}({\bf e}_i)_\alpha({\bf e}_i)_\beta v_\alpha v_\beta-
%\frac{1}{6} v_\alpha v_\alpha\right ], \nonumber\\
%f_0^{r(eq)}&=&\rho_r\left [\frac{m_r}{6+m_r}-v_\alpha v_\alpha\right ],
%\nonumber \\
%f_i^{b(eq)}&=&\rho_b\left [\frac{1}{6+m_b}+\frac{1}{3}({\bf e}_i)_\alpha
%v_\alpha+\frac{2}{3}({\bf e}_i)_\alpha({\bf e}_i)_\beta v_\alpha v_\beta-
%\frac{1}{6} v_\alpha v_\alpha\right ], \nonumber\\
%f_0^{b(eq)}&=&\rho_b\left [\frac{m_b}{6+m_b}-v_\alpha v_\alpha\right ].
f_i^{r(eq)}&=&\rho_r\left [\frac{1}{6+m_r}+\frac{1}{3}({\bf e}_i\cdot{\bf v})
+\frac{2}{3}({\bf e}_i\cdot{\bf v})^2-
\frac{1}{6} v^2\right ], \\
f_0^{r(eq)}&=&\rho_r\left [\frac{m_r}{6+m_r}-v^2\right ], \\
f_i^{b(eq)}&=&\rho_b\left [\frac{1}{6+m_b}+\frac{1}{3}({\bf e}_i\cdot{\bf v})
+\frac{2}{3}({\bf e}_i\cdot {\bf  v})^2-
\frac{1}{6} v^2\right ], \\
f_0^{b(eq)}&=&\rho_b\left [\frac{m_b}{6+m_b}-v^2\right ].
\end{eqnarray*}
Note here that  two real-valued parameters, $m_r$ and $m_b$ have been
introduced, which allow the system to have the density ratio:
 $$\gamma=\frac{\rho_r} {\rho_b} = \frac{6 + m_r}{6 + m_b}.$$

The second part of the collision operator, $(\Omega^{k})^{2}$, was first
 given by Gunstensen {\em et al.}\cite{gun}:
\[ (\Omega_i^{k})^{2} = \frac{A_k}{2}|{\bf F}| (({\bf e}_i \cdot {\bf F})^2
/|{\bf F}|^2 - 1/2), \]
where ${\bf F}$ is the local color gradient, defined as:
\[ {\bf F}({\bf x}) = \sum_i {\bf e}_i (\rho_r({\bf x}+{\bf e}_i) -
\rho_b({\bf x}+{\bf e}_i)). \]
Note that in a single-phase region of the incompressible fluid model,
${\bf F} = 0$.
Thus the second term of the collision operator, $(\Omega_i^{k})^{2}$,
only contributes at two-phase interfaces.  The parameter, $A_k$, is
a free parameter, controlling the surface tension (as shown below).
To maintain interfaces between fluids, the Rothman
scheme is used \cite{roth2} to force
the red color momentum, ${\bf j}^{r} = \sum_{i} f_i^{r} {\bf e}_i$,
to align with the direction of the
local color gradient.  In other words, one can redistribute the red density at
an interface to maximize the following quantity:
$-({\bf j}^{r} \cdot {\bf F}).$
The blue particle distribution can then be obtained using mass conservation
along each direction,
$f_i^{b} = f_i - f_i^{r} (i=0,\cdot,\cdot,\cdot,6)$.

Going through a Chapman-Enskog expansion
procedure similar to Sec. 2, one can rigorously prove that the Navier-Stokes
equation will be valid for each phase with the
additional property that the
pressure difference in the interface will obey
the so called Laplace formula:
\[ \Delta P = P_r - P_b = 2 \sigma / R, \]
where R is the radius of curvature of the surface at the contact point and
$\sigma$ is the
surface tension. Numerical simulations have demonstrated that $\sigma$ is a
constant independent of $R$.

The lattice multiphase models have been used for investigating interface
dynamics, including the Rayleigh-Taylor instability, the Saffman-Taylor
instability and
the domain growth in spinodal decomposition.
The two-phase Rayleigh-Taylor instability analysis predicted
by the lattice method for the
most unstable mode agrees well with
 linear instability theory\cite{roth3}. The numerical prediction from lattice
methods for structure functions and domain
growth with time in two-dimensional and  three-dimensional binary fluids agree
very well with  experiments and theoretical predictions\cite{frank2}. The
two-phase lattice Boltzmann
model has also been extended to three and four phases to study  critical
 and
off-critical quenches\cite{turab}. It has been
 shown that  domain growth
studies agree  well with theoretical predictions\cite{turab}.

Lattice Boltzmann methods have been also used for studying the self-assembly
biological membranes\cite{turab2}. Recently Lookman {\em et al.} at Los Alamos
National Laboratory have developed a scheme which incorporates the
surfactant effects in two- and three-dimensional multiphase models. They have
studied the effects of adhesion, micelles in bio-membranes and self-assembly of
amphiphilic structures at aqueous organic
interfaces. In Fig. 11, we present the two-dimensional lattice Boltzmann
results from one such simulation.

In Fig. 12, we show  a two-dimensional Hele-Shaw viscous fingering
experiment. A no-slip boundary condition is assigned to the upper and lower
walls, and the fluids are given a viscosity ratio of 1:10 by
setting $\tau_r=1$ and $\tau_b=5.5$.  To develop a Hele-Shaw pattern,
an initial perturbation is introduced at the interface\cite{perturb,degregoria}
which is then forced by maintaining a high pressure at the inlet and low
pressure at the outlet.  The shape of the finger can be altered by
changing the surface tension parameter, $A$, which is $0.0065$ in this
experiment.  The plot shows the evolution of a less viscous fluid penetrating
one having a higher viscosity at times = 0, 2000, 4000, 6000, 8000 and 10000.
We see that a stable finger develops and is maintained throughout the
simulation.
The fluid patterns shown here are similar to results obtained by other
numerical methods \cite{degregoria}.

Lattice Boltzmann multiphase fluid models have been used for simulating fluid
flows in porous media for understanding fundamental physics related to
enhanced oil recovery and predicting the
relative permeability. Simulation results compare well with experimental
measurements of flow patterns and relative permeability\cite{daryl2}.
In Fig. 13, we present a three-dimensional snapshot of oil and water
distribution from one of these simulations
which was carried out at Los Alamos National Laboratory using
the Connection Machine 5.  The simulation shown here tests oil
recovery by water flooding.
As in experimental
 enhanced oil recovery, the water (blue) is being used to drive the oil
(red) out of the porous media
(transparent).
As seen in the plot, the lattice Boltzmann simulation preserves the
fundamental phenomena observed in real experiments that the water phase forms
long fingers through the porous  media due to the wettability properties of the
water. A great deal of effort has gone into the studies of the effects of
wettability on relative permeability versus saturation curves. Because of its
ability to represent the exact Navier-Stokes equation in a parallel fashion,
to handle arbitrary complicated geometry and simulate surface dynamics and
wettability, the lattice Boltzmann method is becoming  an increasingly
popular means of modeling multiphase fluid flows in porous media.

\section{Macromolecular Dynamics and Polymeric fluids}

Most lattice methods introduced so far involve particles with
point-like properties of mass, momentum and
energy, but with zero radii and no internal structure.
There are two kinds of non-point particle macromolecular models so far which
extend
the original lattice gas and lattice
Boltzmann models to simulate the
evolution and interaction of extended objects, such as colloidal suspensions
and polymeric liquids.

A simple idea for using lattice gas models to study scaling dynamics in
polymeric fluids
was proposed by Vianney and Koelman\cite{vianney}.
The basic idea in this model is to enforce some constraints on particle free
steaming and to
form  a set of particles (chain).
Each individual particle in the chain
cannot move very far
from the others  if they are in the same set. A one-dimensional
chain can move  in one, two or
three spatial dimensions with some new rules and the particles on a chain
can interact with free particles and other particles in another chain.
One additional feature of this model is that at most
one particle  can be at each lattice node.
This is a more restrictive condition than the exclusion principle
used in lattice gas models with  point-like particles, where as many particles
as nearest neighbors are usually allowed at each grid point.
On the other hand,  particles coming from  various neighboring sites
can aim towards the same grid point during the propagation step  and this
makes the algorithm  more difficult to implement
on a parallel machine. This problem is resolved in Ref. \cite{vianney}
by introducing a ``trial and error'' step, called collision step,
in which all particles that aim at the same grid point change their
velocities several times, until they can all move to a different site.
The change in velocities is done to groups
of particles simultaneously, imposing   conservation of
momentum and kinetic energy at each collision. The computer experiments
analyzed in Ref. \cite{vianney} showed convergence of this procedure
(at least in cases with 25\% of the sites occupied by particles).
After the collision step, the velocities
of the particles that belong to each different
chain are changed so that they will not move too far apart from one
another during the subsequent propagation step.
This  update step  is done by minimizing the length that the chain
would have after propagation, subject to the condition of single occupancy
of each site. After the update step, all particles are moved to their
 neighboring sites according to their corresponding velocities.
All previous steps guarantee that each particle will move to a different
site.

The authors were able to show that the model can be used to study scaling
dynamics for a chain of macromolecular particles moving in a solvent when the
hydrodynamic effect is important and the excluded-volume is imposed.
 They have measured the radius of gyration, $R_g$,  as a function of the number
of
chain particles for different particle velocity and lattice size. The anomalous
scaling exponents agrees well with theoretical predictions.  They have also
found
that the center-of-mass velocity autocorrelation function of a single chain
in a two-dimensional solution obeys a dynamic scaling relation which
violates the so-called ``nondraining'' concept. \cite{vianney}

It appears that  the numerical calculation in this model
becomes more complex with the
 increasing
 number of chain particles.
This actually limits its applicability for real polymeric dynamics.

The model proposed in Ref. \cite{chopstring} appears more realistic,
in the sense that it treats the interaction between particles in a
chain as if they were connected by springs. Each chain is assumed
to be composed of two types of particles, ``white'' and ``black'',
which are distributed along the chain in an alternating fashion.
There is an
even number of ``white'' particles and an odd number of ``black''
ones, a white particle being at the beginning and at the end of
each chain. Each particle has an associated mass, $m_i$, where $i$ labels
the location of the particle in the chain. It is assumed
that all particles, except the ones at the ends, have a mass equal
to one, while the two at the ends have a mass equal to $1/2$.
The propagation step is divided into different substeps, each corresponding
to a movement along a different spatial direction. Furthermore, each
substep is subdivided into a propagation of the white particles
first and then the black ones. As in the previous model, only
one particle of the chain is allowed at each lattice site. However,
the update of the positions on the lattice is done in a completely
 deterministic
way. One of the parameters of the model is the maximum distance
allowed between consequent particles in a chain, which also
determines the spring constants that rule their interaction.
In the case of the longitudinal propagation, we can interpret the
step as governed by the following rule:
if the separation between two particles is large, then the
springs connecting them pulls them together. Therefore,
the particle whose position is being updated
(remember that alternating particles are updated at alternating
times), is moved towards the other one. If
the separation is small, then the spring
pulls them apart, and so, the corresponding particle is moved away
from its neighbor. This model possesses three quantities that are
conserved during the propagation step: total mass, $M\equiv
\sum_{i=0}^{2N} m_i$, total
momentum, ${\bf P}
\equiv\sum_{i=0}^{2N}
 m_i({\bf x}_i(t+1)-{\bf x}_i(t))$ and potential energy, $E
\equiv\sum_{i=0}^{2N-1}\vert{\bf x}_{i+1}(t)-{\bf x}_i(t)-{\bf a}\vert^2$,
where $|{\bf a}|$  is the natural spring length, $\frac{3}{2}$ in the Chopard's
model.
The total momentum is proportional to the velocity of the center
of mass, which is also conserved during the propagation step.
In the one-dimensional case, the energy is not an independent conserved
quantity, but it is proportional to the total mass.
Collisions between chains are not described in the general three-dimensional
case, though it is mentioned that they should maintain the conservation
of mass, momentum and energy. The main difficulties in the implementation
of realistic collisions in three-dimensions are due to the fact that
it would be necessary to check whether the post-collision configuration
is allowed  and if each particle is involved in more than one
collision. Therefore, the author only presents an example in one
dimension that satisfies these conditions. In one dimension the
problems mentioned before do not arise,
since only the particle at the end of a chain
can collide with another end-particle, provided that they have opposite
velocities. Since collisions between chains can be complicated, the
authors suggest that it would be interesting to study, as a first
step, the interaction of a single chain with point-like particles.
This study would provide  insight into the problem of an object in
a non-equilibrium fluid or in other problems of geological interest.
Finally, it is mentioned that it would be important
to extend this model to include  rotations
of the chains, with their associated conserved angular momentum.

The first model using the lattice gas automaton method
 for colloidal suspensions was proposed by Ladd and Frankel\cite{ladd1}.
They considered the moving objects interacting with fluid particles
by extending the usual ``bounce-back'' rule for nonslip
boundary conditions. In order to allow macromolecular particles interaction
with particles when the molecule
 has a nonzero velocity, the inside and outside of the moving objects are both
occupied by  particles. A new microscopic collision
rule was introduced for particles colliding with the
macromolecular surfaces.  The
microrules at the boundary nodes include independent rules for each
pair of velocity directions, $ {\bf e}_i$ and
its opposite direction ${\bf e}_{-i} = - {\bf e}_i$. A small fraction, $p_i$,
of the particles moving in the
$i$ direction are remained unperturbed, while the remainder of these particles
and all those moving in the $i$ direction are reflected. The parameter,
$p_i$, is chosen so that the particles at the node have the
same mean velocity as the
speed of the macromolecular
interface. The momentum flux of the particles can be used to calculate
the translation and rotation speed of the macromolecular
particles in the next step. This simple lattice gas model
for moving objects has
achieved considerable
 success in simulating suspension particles in fluids and agrees
well with experimental measurements and  theoretical predictions,
including the calculation of the drag coefficient of  a
spherical particle moving in fluids, the long-time tails of the
 translation and rotation correlation functions of a colloidal suspension
in
two-dimensional fluids.

The above lattice gas model of moving object in fluids has been extended
by Ladd\cite{ladd3} and Chen {\em et al.} \cite{chenball} recently
 in order to have correct hydrodynamics
and to reduce the noise in the lattice gas models. The
work by Ladd
is basically the same as
his lattice gas model. Since the particle fluctuations
are smoothed out in the lattice Boltzmann scheme, Ladd, for the first time,
introduces a random fluctuation term
in the lattice Boltzmann equation in order to account for the
fluctuation-dissipation mechanism of Brownian motion.
 The new kinetic equation with fluctuations can be written as follows:
\begin{equation}
f_i({\bf x}+ {\bf e}_i,t+1)=f_i({\bf x},t)+\Omega_i (f({\bf x},t)) + f'_{i},
\end{equation}
where $f'_i$ is chosen so that the  additional stress tensor,
$\Pi' = \sum f'_{i} {\bf e}_i{\bf e}_i$,
 is nonzero and the random
stress fluctuations are uncorrelated in space and time,
\[ \langle \Pi_{\alpha \beta}({\bf x}, t)\Pi'_{\gamma \delta}({\bf x'},
t')\rangle
= A \delta_{{\bf x} {\bf x'}}\delta_{t t'}(\delta_{\alpha \gamma}
\delta_{\beta \delta} + \delta_{\alpha \delta}\delta_{\beta \gamma} -
\frac{2}{D} \delta_{\alpha \beta}\delta_{\gamma \delta})
. \]
The coefficient, A, is related to
temperature. Using his new lattice Boltzmann
model, Ladd simulated the dynamics of short time motion of colloidal
particles. The results compare favorably with diffusing-wave spectroscopy
experiments.

Following the Ladd and Frankel scheme of allowing particles to move inside
macromolecular particles in their lattice Boltzmann methods,
Chen {\em et al.}\cite{chenball} developed a new method for implementing the
interaction between fluid particles and macromolecular particles.
In the Ladd and Frankel
 lattice gas method, particles bounce
back from a
wall for zero wall velocity.
The lattice Boltzmann
method by Ladd is basically an exact copy from the lattice gas scheme, simply
exchanging
 particle bounce-back  for
distribution function
bouncing back. Chen {\em et al.} recognized the difference between
the  lattice gas  and the
lattice Boltzmann methods. Instead of following the particle picture, they
let the particle distribution function at the surface at each time step to
satisfy the non-slip condition, i.e. they require
the distribution
function of particles to have the same velocity as the macromolecular
particles.
By doing this, they calculate the momentum flux directly without following each
point on the boundary surface. The idea for the latter model appears to be
simpler
and the computational speed in a parallel machine is
 faster than the original scheme.

\section{Concluding Remarks}

In this paper, we have introduced the basic principles of lattice gas and
lattice Boltzmann methods and described several important extensions, including
lattice models for reaction-diffusion systems, chemical precipitation
and dissolution at a
 surface, multiphase fluid flows in porous media and polymeric liquids. In
general, simulation results obtained by lattice methods are in
 good quantitative
agreement with experimental results and simulation results from other numerical
methods. Lattice methods
are useful numerical modeling tools for
problems related to chemically reacting systems.

It should be mentioned that lattice methods are a
mesoscopic dynamic description of physical phenomena.  Problems for which
macroscopic hydrodynamics and microscopic statistics are both important at the
same time can be simulated. For example,
in this paper we have presented the noise-induced pattern  formation in a
reaction-diffusion system where the microscopic fluctuation and  macroscopic
partial differential equations both contribute to the dynamics. The intrinsic
fluctuations allow us to follow the nonequilibrium dynamics, while their
averaged dynamics satisfy the  reaction-diffusion equations.

The lattice automaton method
uses bit operations and is unconditionally stable. On the other hand,
the lattice Boltzmann
method is
 similar to an explicit finite difference scheme, solving the
discretized kinetic equation by
a Lagrangian upwind scheme\cite{sterling,maria2,peter}, which is
  conditionally stable. Stability and accuracy analyses have been provided
recently by Sterling and Chen\cite{sterling}, and
by Reider and Sterling\cite{reider}, respectively. It has been shown from their
analyses that the lattice Boltzmann
methods presented in this paper are
second order accurate  both in
time and space. To ensure numerical stability, the relaxation time should
not be less than $\frac{1}{2}$. (See details in \cite{sterling} for other
parameter dependencies of the
stability of lattice Boltzmann methods).

Lattice methods are ideally suited to implementation on parallel
computers. Not only is the scheme more
efficient  conputationally than
 traditional numerical schemes,
but it is also more efficient in the sense that
 the time period for producing computational results is much shorter and
the work required to develop
 codes is less than that for other
numerical codes.

Lattice methods are still in the development stage.
Though the fundamental theory and principles are
solidly
based on statistical mechanics,
there remain  several unsolved problems and developing areas.
The current lattice models for reacting systems are based on
simple isothermal models, but this can easily be extended
to include  temperature
effects and to
model
 phase-transitions and allow the reaction rate to
depend on local temperature.

\section{Acknowledgements}

We thank F. Alexander, M. Ancona, M. Dembo,
D. Grunau, K. Eggert, B. Hasslacher,
 S. Hou, R. Kapral, A. Lapedes, T. Lookman, L. Luo,
D. Martinez, W. Matthaeus, B. Nadiga, J. Pearson and J. Sterling for useful
discussions.
This work is supported by the US Department of Energy at
Los Alamos National Laboratory. Numerical
simulations were done on the CM-5 at the Advanced Computing Laboratory
at Los Alamos National Laboratory.

\vfill\eject

\section{Figure Captions}
\begin{description}

\item[Fig. 1]Three simple collision rules in the lattice gas automata: two-body
head-on collision, three-body symmetric collision and one moving and one rest
particle collision.  The left-hand side configurations
represent the initial states
and the right-hand configurations are the final states.

\item[Fig. 2] Vorticity distribution in Kelvin-Helmholtz instability
using the lattice gas  method
for time = 0,  15000, 20000, 120000 and 490000.
One can follow the pairing process
for similar sign vortices.

\item[Fig. 3] Comparisons between the spectral method (SP)
and the lattice Boltzmann (LB) simulations for the velocity contour
lines at ($z = 64$). The left plots are at $t = 4$ and the right plots
are at $t= 10$.  The similar velocity structures for different time
steps can be clearly seen from the plots.  This demonstrates that the two
numerical schemes agree well not only globally, but also locally\cite{sycjsp}.

\item[Fig. 4]Vorticity Distribution for flow around a blunt body by lattice
Boltzmann simulation done by L. Luo\cite{luo1}. The Reynolds number is about
1000. The simulation reproduces the attached vorticies and
vortex shedding phenomena.

\item[Fig. 5]
 Spontaneous nucleation of rings of excitation in the
excitable region and spiral wave formation.  In addition to ring
growth from a disc-shaped perturbation, followed by shearing of the
ring and spiral wave formation, waves of excitation spontaneously
nucleate.  In the last frame one sees that the spiral wavefronts have
collided and ``pinched off" a wave fragment that can serve as the nucleus
of another spiral wave pair.
System parameters:
$k_1\rho_A=0.00008385$, $k_{-1}=0.1k_3$, $k_3=0.0005$, $k_2=k_{-2}=0.01$
and $k_{-3}\rho_B=0.000002326$ and the system size is $1024 \times 1024$.
The propagation and rotation steps were performed
every three
chemical reaction steps\cite{refkap1}.

\item[Fig. 6]
Reactive lattice gas automaton Turing structures.  The
Turing pattern  evolved from a random initial condition.   The microscopic
 dynamics was constructed (Ref. \cite{refkap1}) to correspond to
the Sel'kov rate law with the following kinetic parameters:
$k_1\rho_A=0.002656673$, $k_3=0.00665=10k_{-1}$, $k_2=k_{-2}=0.015$ and
$k_{-3}\rho_B=0.000531334$.  The concentration of the product, $Y$ is
displayed.
  The diffusion
coefficients are: $D_X=50D$ and $D_Y=2D$ with $D=0.25$ in units of the lattice
spacing squared per time step.  The simulation was carried out on a square
lattice ($512 \times 512$) with periodic boundary condition.  The four panels
(left to right and top to bottom) are for
times $t=0$, $t=2000$, $t=10000$ and $t=20000$\cite{refbrosl}.

\item[Fig. 7]  Spatial distribution of the density,  $n_2$, of species $Y$
at times $t=4,000$, $t=8,000$ and $t=50,000$
for a pure reaction-diffusion system with Sel'kov kinetics (\ref{eq:selkov})
and $k_1\rho_a=0.002656673$, $k_3=0.00665=10k_{-1}$, $k_2=k_{-2}=0.015$,
$k_{-3}\rho_b=0.000531334$,  $D_1=0.2914$ and
$ D_2=0.0170$
 (upper row)
and  at times  $t=4,000$, $t=8,000$ and $t=16,000$
for the same situation but in the presence of a moving solvent
that flows with velocity ${\bf v}({\bf x},t) =  v_x (y) \hat x$\cite{refsil}.
 Blue, green, yellow and red
colors represent increasing values of density concentration.
We can observe that, in the absence of the moving solvent,
an hexagonal pattern is formed, while when the flow is introduced,
the rotational symmetry is broken and the final pattern is
a set of stripes.

\item [Fig. 8] Comparison of lattice gas (upper row) and
lattice Boltzmann (lower row) simulations of pattern
 formation in a reaction-diffusion system
with Sel'kov kinetics (\ref{eq:selkov})
for which
 $k_1\rho_a=0.005$, $k_{-1}=0.001$, $k_2=k_{-2}=0.002$, $k_3=0.00395$ and
$k_{-3}\rho_b=0.000005$ and the diffusion coefficients are  $D_1=1$ and
$D_2=0.125$ for the lattice gas and $D_1=1.0286$ and
$D_2 = 0.1286$ for the lattice Boltzmann scheme.
 A spot of high  $Y$ concentration  is initially slightly
deformed and finally splits into two new spots of high $Y$
concentration. The internal fluctuations in the lattice gas simulation
are evident.

\item [Fig. 9]Snapshot of chemical evolution of a simple pore network in a
lattice gas model with one component
solid in which zero concentration solution is input and walls
dissolve as a function of local concentration utilizing
transition-state
theory rate law\cite{wells2}. Color code corresponds to concentration.

\item [Fig. 10]Transport through a complex,
heterogeneous network (natural rock pore structure digitized into
$512 \times 256$ grid) in a lattice Boltzmann model
 for square-wave injection of
non-sorbing
chemical tracers.  Color code corresponds to log concentration scale
of six orders of magnitude with maximum of 20 ppm concentration.

\item [Fig. 11]Two dimensional bio-continuum phase by the lattice Boltzmann
simulation. Two phase are represented by red and
 dark blue colors. The light blue
color represents the surfactant.

\item[Fig. 12]The formation of a stable finger in a Hele-Shaw cell at times
$t =$ 0, 2000, 4000, 6000, 8000 and 10000.  A viscosity ratio of
1:10 is established by setting $\tau_r=1$ and $\tau_b=5.5$.  The
low-viscosity fluid  is penetrating the higher viscosity region\cite{daryl}.

\item[Fig. 13]The three-dimensional snapshot of oil and water distribution
from the lattice Boltzmann simulation. The red color represents the
non-wetting
fluid (oil), the blue color represents the wetting phase (brine) and the rock
is transparent.
\end{description}

\end{document}